\documentclass{aastex62}
\usepackage{amsmath,amssymb,latexsym} 
\usepackage{xcolor, fontawesome} 
\usepackage{graphicx}
\usepackage{lipsum}

\begin{document}

\title{Carbon-Oxygen Classical Novae are Galactic $^7$Li Producers as well as Potential Supernova Ia Progenitors}

\author[0000-0002-1359-6312]{Sumner Starrfield}
\affiliation{Earth and Space Exploration, Arizona State University, P.O. Box 871404, Tempe, Arizona, 85287-1404, USA
starrfield@asu.edu}

\author[0000-0002-7978-6570]{Maitrayee Bose}
\affiliation{Earth and Space Exploration, Arizona State University, P.O. Box 871404, Tempe, Arizona, 85287-1404, USA}
\affiliation{Center for Isotope Analysis (CIA), Arizona State University, Tempe, Arizona, 85287-1404, USA}

\author[0000-0003-2381-0412]{Christian Iliadis}
\affiliation{Department of Physics \& Astronomy, University of North Carolina, Chapel Hill, NC 27599-3255}
\affiliation{Triangle Universities Nuclear Laboratory, Durham, NC 27708-0308, USA}

\author[0000-0002-9481-9126]{W. Raphael Hix}
\affiliation{Physics Division, Oak Ridge National Laboratory, Oak Ridge TN, 37831-6354}
\affiliation{Department of Physics and Astronomy, University of Tennessee, Knoxville, TN 37996 }

\author[0000-0001-6567-627X]{Charles E. Woodward}
\affiliation{MN Institute for Astrophysics, 116 Church Street, SE University of Minnesota, Minneapolis, MN 55455 }

\author[0000-0003-1892-2751]{R. Mark Wagner}
\affiliation{Large Binocular Telescope Observatory, Tucson, AZ 85721}
\affiliation{Department of Astronomy, Ohio State University, Columbus, OH 43210}

\newcommand{\kms}{km\,s$^{-1}$}
\def\lesssim{\mathrel{\hbox{\rlap{\hbox{\lower4pt\hbox{$\sim$}}}\hbox{$<$}}}}
\def\gtrsim{\mathrel{\hbox{\rlap{\hbox{\lower4pt\hbox{$\sim$}}}\hbox{$>$}}}}
\def\apj{$Astrophys.\ J.$}
\def\apjl{$Astrophys.\ J.$}
\def\aj{$Astron.\ J.$}
\def\aap{$Astron.\ Astrophys.$}
\def\mnras{$Mon.\ Not.\ R.\ Astron.\ Soc.$}
\def\pasj{$Publ.\ Astron.\ Soc.\ Jpn$}
\def\iaucirc{$IAU Circ.$}


\begin{abstract}
We report on studies of Classical Nova (CN) explosions where we follow the evolution of thermonuclear runaways (TNRs) on Carbon Oxygen (CO) white dwarfs (WDs). 
We vary both the mass of the WD (from 0.6 M$_\odot$ to 1.35 M$_\odot$) and the composition of the accreted material. 
Our simulations are guided by the results of multi-dimensional studies of TNRs in WDs that find 
sufficient mixing with WD core material occurs after the TNR is well underway, reaching levels of enrichment that 
agree with observations of CN ejecta abundances. We use NOVA  (our 1-dimensional hydrodynamic code) to accrete
solar matter until the TNR is ongoing and then switch to a mixed composition (either 
25\% WD material and 75\% solar or 50\% WD material and 50\% solar).  
Because the amount of accreted material is inversely proportional to the initial $^{12}$C abundance, by first accreting 
solar matter the amount of material taking part in the outburst is larger than in those simulations where we assume a mixed composition from the beginning. 
Our results show large enrichments of $^7$Be in the ejected gases implying that CO CNe may be responsible
for a significant fraction ($\sim$ 100 M$_\odot$) of the $^7$Li in the galaxy ($\sim$1000~M$_\odot$).  
In addition, although the ejected gases are enriched in WD material, the WDs in these simulations eject less material than they accrete.  We predict that
the WD is growing in mass as a consequence of the accretion-outburst-accretion cycle and CO CNe may be an important
channel of Supernova Ia progenitors.  
 \end{abstract}

\section{Introduction}

Classical Novae occur in close binary systems with a white dwarf (WD) primary and a secondary which is a larger
cooler star that fills its Roche Lobe.  It is losing material through the inner Lagrangian point which ultimately is
accreted by the WD.  These binary systems are referred to as Cataclysmic
Variables (CVs).  The consequence of the WD accreting sufficient material is
a thermonuclear runaway (TNR) in matter that is electron degenerate at the beginning of accretion and
thus produces an event that is designated a ``nova outburst'' (either Classical, Recurrent, or Symbiotic Nova; hereafter CN, RN, or SymN). 
While the observed outburst ejects material into the surrounding region, 
it does not disrupt the WD and continued
accretion implies successive outbursts.  In some cases, the properties of the WD and accretion result in
outbursts repeated on human time-scales which are designated RNe.  If the orbital separation is large and the
secondary is a red giant, then the system is designated a SymN.   

The observations of the chemical composition of the gases ejected by a CN explosion, show that they typically are 
extremely non-solar \citep{warner_1995_aa, gehrz_1998_aa, bode_evans_08, starrfield_2012_basi, downen_2012_aa}. 
Because of the CNe observations, 
it is assumed that the accreting material mixes with the outer layers of the 
WD at some time during the evolution from the beginning of accretion 
to the observed outburst.  Thus, the observed ejected gases consist of a 
mixture of WD and accreted material that has been processed by hot-hydrogen burning.  It has also been assumed 
that the CN outburst ejects more mass (both accreted and WD matter) from the WD than accreted from the secondary star
and, therefore, the WD is decreasing in mass as a result of continued CN outbursts and it cannot be a progenitor of Supernova of type Ia (SN Ia).  
In contrast, if the WD accretes more mass than it ejects during the CN outburst,
then it is growing in mass and could possibly reach the Chandrasekhar Limit and explode
as a SN Ia.  In this paper, we report on our new simulations of the CN outburst and find that
the WD is ejecting less mass than accreted and, therefore, the WD is growing in mass
and CO CNe could be one of the channels for the progenitors of SN Ia explosions.

SN Ia are the optically brightest explosions that occur in a
galaxy and they can be detected to, at least, z $\sim$ 2 in the universe.  Studies of SN Ia
show that their light curves are standardizable, allowing them to be used as distance indicators, 
which led to the discovery of dark energy in the universe \citep{riess_1998_aa, perlmutter_1999_aa}.  In addition, they contribute a 
significant fraction of the iron-group elements to the galaxy and the solar system \citep{hillebrandt_2000_aa,leibundgut_2000_aa}.  However,
the systems that actually explode as a SN Ia are as yet unknown.  Two pathways are currently posited, 
the single-degenerate (SD) and the double degenerate (DD).  The
DD scenario requires either the merging or collision of two
carbon-oxygen (CO) WDs while the SD scenario assumes
that a CO WD exists in a close binary stellar system and it is growing in mass
toward the Chandrasekhar Limit \citep{hillebrandt_2000_aa,leibundgut_2000_aa, leibundgut_2001_aa, maoz_2014_aa, ruiz_2014_aa, nugent_2019_aa}. 
Therefore, the determination of the response of a CO WD to the CNe phenomenon (growing or shrinking in mass) may shed light on 
one channel of SN Ia progenitors.

Another important motivation for studies of the consequences of TNRs on CO WDs is the recent discovery of both $^7$Li and $^7$Be in
the early high dispersion optical spectra of the ejected material from CN outbursts  
\citep{tajitsu_2015_aa, tajitsu_2016_aa, izzo_2015_aa, izzo_2018_aa, molaro_2016_aa, selvelli_2018_aa,wagner_2018_aa} 
which has validated earlier predictions \citep{starrfield_1978_aa,
hernanz_1996_aa, jose_1998_aa, yaron_2005_aa} and warrants new theoretical studies. 
CNe produce $^7$Li via a process originally described by \citet{cameron_1971_aa} for red giants.   
\citet{starrfield_1978_aa} then applied their mechanism to CN explosions, but that version of NOVA did not, as yet,
include accretion and they assumed that the envelope was already in place.  Later
\citet{hernanz_1996_aa} and \citet{jose_1998_aa} followed the accreting material and were able to investigate
the  formation of $^7$Be during the TNR.  They determined the amount of $^7$Be
carried to the surface by convection and surviving before it could be destroyed by the $^7$Be(p,$\gamma$)$^8$B
reaction occurring in the nuclear burning region.  If it survives by being transported to cooler regions, $^7$Be decays via electron-capture to $^7$Li with an $\sim$
53 day half-life \citep{bahcall_1969_ab} .  

The studies reported in this paper confirm that a TNR on a CO WD overproduces $^7$Be with
respect to solar material and in amounts that imply that such CNe are responsible for a significant amount of galactic $^7$Li.
In contrast, $^6$Li is produced by spallation in the interstellar medium \citep{fields_2011_aa} and its abundance in the solar system should not
correlate with $^7$Li.  \citet{hernanz_2015_aa} gives an excellent discussion of the cosmological importance of detecting $^7$Li in nova explosions. 
We return to this comparison in Section \ref{lithium}.

Here, we investigate both the SD scenario and the production of $^7$Li in the CN outburst by simulating accretion onto CO
WDs, in which we include mixing of the WD outer layers with accreted solar matter 
after the TNR has been initiated.  
We report on three separate studies. First, we accrete mixed material (either 25\% WD matter and 75\% solar matter or
50\% WD matter and 50\% solar matter) from the beginning of accretion.  This is the procedure used in the past for both
accretion onto CO WDs and ONe WDs \citep[][and references therein]{starrfield_1997_aa, jose_1998_aa, starrfield_2009_aa, hillman_2014_aa, starrfield_2016_aa}.  
However, we find, as reported later, that these explosions do not agree with the observed aspects of CNe
outbursts \citep{warner_1995_aa, starrfield_2012_basi, bode_evans_08}.  
Second, we accrete solar matter from the beginning and follow the resulting evolution through the peak and the
return to nuclear burning quiescence.  Third, we take the solar accretion simulations and once the TNR is ongoing (T $ \sim 7 \times 10^7$K),
we stop the evolution and switch the accreted layers to either of the mixed compositions noted above.
This last set of simulations is guided by the results of multidimensional (Multi-D) studies of mixing on WDs which indicate
that sufficient material is dredged up from the outer layers of the WD {\it during the TNR} to agree with observed abundances \citep{casanova_2011_ab, casanova_2016_aa, 
casanova_2018_aa, jose_2014_aa}.

In the next section we discuss NOVA our 1-D hydrodynamic computer code. We follow that with the sections where we present the
results for each of the above studies and end with a discussion and conclusions. 

\section{NOVA: Our 1-Dimensional Hydrodynamic Code}

We use NOVA \citep{kutter_1972_aa,sparks_1972_aa, kutter_1974_aa, kutter_1980_aa, starrfield_2009_aa, 
starrfield_2016_aa} in this study.   The most recent description of NOVA can be found in 
\citet[][and references therein]{starrfield_2009_aa}.  NOVA is a  one-dimensional (1-D), fully implicit, hydrodynamic, 
computer code that has been well tested against standard problems \citep{kutter_1972_aa, sparks_1972_aa}.
NOVA  includes a large nuclear reaction network that includes 187 nuclei (up to $^{64}$Ge and including the $pep$ reaction), the OPAL opacities 
 \citep{iglesias_1996_aa},
the Starlib nuclear reaction rates  \citep{sallaska_2013_aa},
the Timmes equations of state \citep{timmes_1999_aa, timmes_2000_ab}, and the nuclear reaction network solver developed by \citet{hix_1999_aa}.   
NOVA also includes the \citet{arnett_2010_aa} algorithm for mixing-length convection and the Potekhin electron degenerate conductivities 
described in \citet{cassisi_2007_aa}.   These improvements have 
had the effect of changing the initial structures of the WDs so that they have smaller radii and, thereby, larger
surface gravities compared to our previous studies.  Finally, we also now 
include the possible effects of a binary companion (an extra source of heating at radii of $\sim 10^{11}$ cm) as described by \citet{macdonald_1980_aa}
which can increase the amount of mass lost during the last stages of the outburst.

In this study, we accreted material at a rate of $1.6 \times 10^{-10}$ M$_\odot$yr$^{-1}$ onto complete CO WDs (the structure extends to the WD center) with 
masses of 0.6 M$_\odot$, 0.8 M$_\odot$, 1.00 M$_\odot$, 1.15 M$_\odot$, 1.25 M$_\odot$, and 1.35 M$_\odot$.  We chose this value of \.M because it is
the value used by \citet{hernanz_1996_aa, jose_1998_aa}, and \citet{rukeya_2017_aa} and later we compare our results to their results.  It is also the value used in our study of accretion
onto ONe WDs \citep{starrfield_2009_aa}. The assumed composition 
of the WD outer layers was 50\% $^{12}$C and 50\% $^{16}$O.  Since $^{12}$C $>$ $^{16}$O and the ratio varies with depth, this can only be considered an approximate value
\citep{althaus_2010_ab, jose_2016_aa, giammichele_2018_aa}.  In fact, it is the amount of $^{12}$C that strongly affects the evolution and
not the C/O ratio.  The basic properties of each WD initial model (luminosity, radius, and
effective temperature) are given in the first 3 rows of Tables \ref{evolCOMFB} and 
\ref{evolCOMDTNR}.  In contrast to our previous studies, we use 150 mass zones with the zone mass decreasing from the center to the surface. 
The mass of the surface zone is $\sim 2 \times 10^{-9}$ in units of the WD mass.  
This is much less than either the accreted mass or the amount of core material mixed up into the 
envelope.  This low a mass decreases the maximum time step  during the accretion phase (which although implicit is
tied to the mass of the outer zone), but allows us to fully resolve the behavior of the simulations as the TNR occurs.  

NOVA follows accretion through the peak of the TNR and the following decline in the temperature toward quiescence.  It allows
us to evolve the expanding outer layers and determine if they are ejected.   
We tabulate, as the ejected mass, the amount of material that is expanding both at speeds above the escape velocity
and also has become optically thin.  We do not remove any mass zones during the evolution as this reduces the numerical pressure
on the zones below causing them to accelerate outward and also reach escape speeds.
We find that even if the material is 
ejected, we can follow the mass zones until they have reached radii of  a few times $10^{12}$cm.  At these radii the density in
the outer layers has fallen to values that are now below the lower limit of the physics (opacity, pressure equation of state,
energy equation of state) tables ($\rho < 10^{-12}$ gm cm$^{-3}$) 
and we end the evolution.

Finally, although it is commonly assumed that a CO WD should not have a mass exceeding $\sim 1.15 $M$_\odot$ \citep{iben_1991_aa, ritossa_1996_aa, iben_1997_aa}, 
as we report in this paper our simulations suggest that WDs are growing in mass,
so that there should be massive CO WDs in CN systems.  An example of this class is Nova LMC 1991, a CO nova,
which exhibited a super Eddington luminosity for more than 2 weeks  \citep{schwarz_2001_aa} likely requiring  a WD mass exceeding $\simeq 1.35$M$_\odot$.  
Moreover, the WDs in four of the nearest CVs
(U Gem: 1.2 M$_\odot$ \citep[][]{Echevarria_2007_aa}, 
SS Cyg: 0.8 M$_\odot$ \citep[][]{Sion_2010_SSCyg_aa}, 
IP Peg: 1.16 M$_\odot$ \citep[][]{copperwheat_2010_aa}, and 
Z Cam: 0.99 M$_\odot$ \citep[][]{Shafter_1983_aa}) are more massive than the canonical value for single WDs of 0.6 M$_\odot$ \citep{sion_1986_aa}.   
More recently, \citet{sion_2018_aa} report a WD mass for the RN CI Aql of 0.98 M$_\odot$, \citet{shara_2018_ab} report that the mean WD mass for
82 Galactic CNe is 1.13 M$_\odot$ and 10 RNe is 1.31 M$_\odot$, while \citet{selvelli_2019_aa} analyzed
18 old novae, using data from both IUE and Gaia, and report that many WDs in CNe have masses above the canonical value for single WDs. 

\section{Simulations with a mixed composition from the beginning}
\label{MFB}

The principal motivation for this paper is to present a new set of simulations where we do not assume a mixed composition until
the TNR is well underway.  However, in order to demonstrate that there is a need for this technique, we first present the 
results of new simulations where we ``Mix From the Beginning'' (hereafter, MFB) as has been done
in nova simulations for many years \citep[][and references therein]{starrfield_2016_aa}. This technique was used because 
there was no consensus on how WD core matter was mixed into the accreted envelope although the observations of
both fast CO and ONe CNe required that such mixing occur \citep[][and references therein]{gehrz_1998_aa, downen_2012_aa, starrfield_2016_aa}.
Nevertheless, it is not physically reasonable to assume the accreted matter is fully mixed from the
beginning of accretion.  A discussion of mixing mechanisms can be found in \cite{jose_2007_aa}.

We find that none of these MFB simulations eject
sufficient material to agree with the observations and for those that do eject some material, the expanding gases have too low a velocity.
This same result was also found in an earlier study of accretion onto CO WDs \citep{starrfield_1997_aa}.  In contrast, our previous ONe CNe simulations \citep{starrfield_2009_aa},
which used the MFB technique, 
did eject significant material because they were initiated with a far lower value for the initial abundance of
$^{12}$C and more material was accreted prior to the TNR than in the CO simulations.  

\begin{figure}[htb!]
\includegraphics[width=1.0\textwidth]{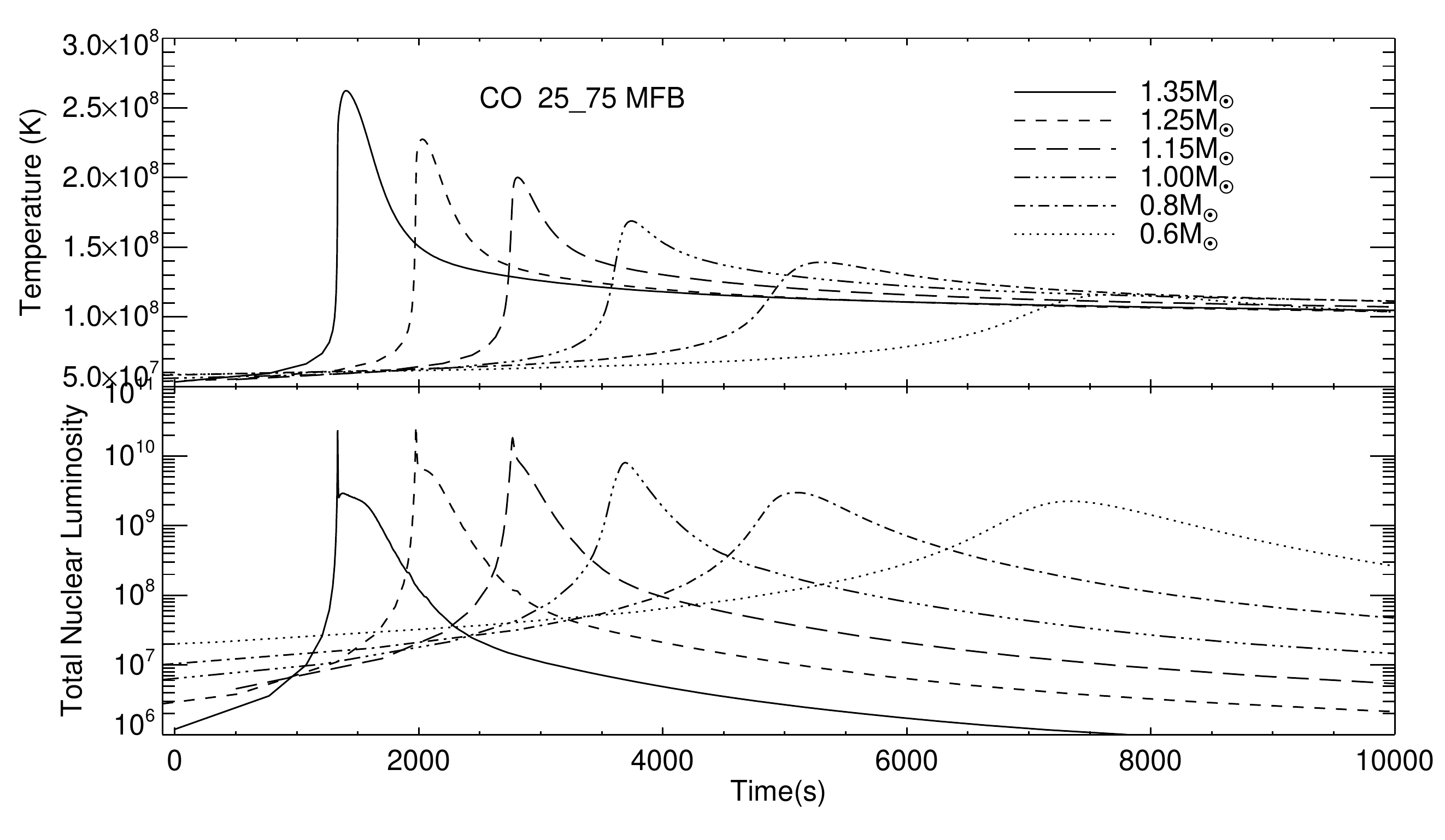}
\caption{Top panel:  the variation with time of the temperature in those mass zones near the interface
between the outer layers of the CO WD and the accreted plus WD matter 
for the simulations with 25\% WD material and 75\% solar material.  In
these simulations we accreted the mixed composition from the beginning (MFB).
The results for all six simulations are shown (the WD mass is identified in
the legend).  The curve for each sequence has
been shifted in time to improve its visibility. As expected, the peak temperature
achieved in each simulation is an increasing function of WD mass.
Bottom panel: The variation with time of the total nuclear
luminosity (erg s$^{-1}$) in solar units (L$_\odot$) around the
time of peak temperature during the TNR. We
integrated over all zones with ongoing nuclear fusion to obtain the plotted numbers. The
identification with each WD mass is given on the plot and the evolution time has again been
shifted to improve visibity.  The cause of the sharp spike at the peak of the curve is discussed in the text.}
\label{MFB2575multi}
\end{figure} 

We use two different mixed compositions in this study.  The first is what we
used in \citet{starrfield_1997_aa} and is 50\% WD matter and 50\% solar matter \citep{lodders_2003_aa}.   The second
composition is 25\% WD matter and 75\% solar in order to better compare our results with \citet{hernanz_1996_aa}, \citet{jose_1998_aa}, and \citet{rukeya_2017_aa} 
who also investigated the consequences of 25\% WD
matter and 75\% solar matter.  In addition, \citet{kelly_2013_aa} studied abundances in ONe novae and reported that the 25\% WD - 75\% solar mixture
was a better fit to the observations.

The initial conditions and evolutionary results for the MFB simulations are given in Table \ref{evolCOMFB}. 
The columns are the values for each of the CO WD masses listed in the top row in solar mass. The first rows give the
initial luminosity, radius, central temperature, central density, and effective temperature for each of the WD masses prior to accretion. As expected,
as the WD mass increases, its radius decreases which is a result of electron degeneracy.  We choose an initial luminosity
of $\sim 4 \times 10^{-3}$ L$_\odot$ in order to obtain as large an amount of accreted mass as possible.  This luminosity
is the same value that we used in our ONe study \citep{starrfield_2009_aa} and only slightly smaller than the $10^{-2}$ L$_\odot$ used in
\citet{jose_1998_aa}.  Increasing the initial luminosity does not change our conclusions; changing the initial chemical composition
has a much larger effect.  Since it is virtually the same initial luminosity for all the WD masses,
as the radius decreases the initial T$_{\rm eff}$ must increase.  
The decrease in radius, in turn, increases the gravitational potential energy at the surface and the
TNR is reached with a smaller amount of accreted mass and, thereby, a smaller accretion time.

\begin{deluxetable}{@{}lcccccc}
\tablecaption{Initial Parameters and Evolutionary Results for Accretion onto CO WDs: Mixing From Beginning (MFB) \label{evolCOMFB}}
\tablewidth{0pt}
\tablecolumns{7}
\tablehead{ \colhead{CO WD Mass (M$_\odot$): }&
\colhead{0.6}&
\colhead{0.8} &
\colhead{1.0} &
\colhead{1.15} &
\colhead{1.25} &
\colhead{1.35}} 

\startdata
Initial:  L/L$_\odot$($10^{-3}$)&4.8&4.8&4.9&4.7&4.8&5.5\\
Initial:  R($10^3$km)&8.5&6.8&5.3&4.2&3.4&2.3\\
Initial: T$_c$($10^7$K)&1.9&1.7&1.6&1.5&1.5&1.5\\
Initial: $\rho_c$ ($10^7$ gm cm$^{-3}$)&$0.34$&$0.95$&$2.9$&$8.3$&$21.0$&$87.0$\\
Initial:  T$_{\rm eff}$($10^4$K)&1.4&1.5&1.7&2.0&2.2&2.7\\
 \hline
{\bf 25\% White Dwarf - 75\% Solar}&&&&&&\\
\hline
 $\tau$$_{\rm acc}$($10^5$ yr)&9.8&3.8&2.0&1.0&0.6&0.2\\
M$_{\rm acc}$($10^{-5}$M$_{\odot}$)&15.5&6.0&3.1&1.7&0.9&0.3\\
T$_{\rm peak}$($10^8$K)&1.2&1.4&1.7&2.0&2.3&2.6\\
$\epsilon_{\rm nuc-peak}$($10^{16}$erg gm$^{-1}$s$^{-1}$)&$0.014$&$0.041$&$0.20$&$0.89$&$2.4$&$6.1$\\
L$_{\rm peak}$/L$_\odot$ ($10^4$)&4.6&4.6&4.4&7.7&4.8&7.0\\
T$_{\rm eff-peak}$($10^5$K)&1.1&3.0&3.4&5.7&8.1&11.0 \\
M$_{\rm ej}$($10^{-7}$M$_{\odot}$)&$8.0$&$3.7$&$0.15$&$0.98$&$0.33$&$0.62$\\
N($^7$Li/H)$_{\rm ej}$/N($^7$Li/H)$_{\odot}$ &$22.0$&$1.1\times10^2$&$1.8\times10^{2}$&$7.9\times10^{2}$&$1.4\times10^{3}$&$2.6\times10^{3}$\\
M$_{\rm ej}$/M$_{\rm acc}$(\%)&0.5&0.6&0.05&0.6&0.4&2.0\\
V$_{\rm max}$($10^2$km s$^{-1}$)&3.9&3.6&4.1&4.6&4.3&5.7\\
\hline
{\bf 50\% White Dwarf - 50\% Solar}&&&&&&\\
\hline
 $\tau$$_{\rm acc}$($10^5$ yr)&6.1&3.2&1.6&0.8&0.5&0.2\\
M$_{\rm acc}$($10^{-5}$M$_{\odot}$)&9.7&5.0&2.5&1.3&0.7&0.2\\
T$_{\rm peak}$($10^8$K)&1.1&1.4&1.7&2.0&2.2&2.6\\
$\epsilon_{\rm nuc-peak}$($10^{16}$erg gm$^{-1}$s$^{-1}$)&$0.015$&$0.081$&$0.33$&$1.4$&$4.3$&$17.0$\\
L$_{\rm peak}$/L$_\odot$ ($10^4$)&2.6&7.3&3.2&8.4&8.1&11.5\\
T$_{\rm eff-peak}$($10^5$K)&2.0&3.6&4.4&6.5&8.6&11.7 \\
M$_{\rm ej}$($10^{-7}$M$_{\odot}$)&$16.0$&$4.1$&$0.44$&$1.3$&$0.83$&$4.0$\\
N($^7$Li/H)$_{\rm ej}$/N($^7$Li/H)$_{\odot}$ &44.0&$1.4\times10^{2}$&$2.9\times10^{2}$&$9.6\times10^{2}$&$1.4\times10^{3}$&$4.3\times10^{3}$\\
M$_{\rm ej}$/M$_{\rm acc}$(\%)&1.6&0.8&0.2&1.0&1.2&20\\
V$_{\rm max}$($10^2$km s$^{-1}$)&0.7&3.6&5.9&12.9&14.9&19.4\\
\enddata
\end{deluxetable}

\begin{figure}[htb!]
\includegraphics[width=1.0\textwidth]{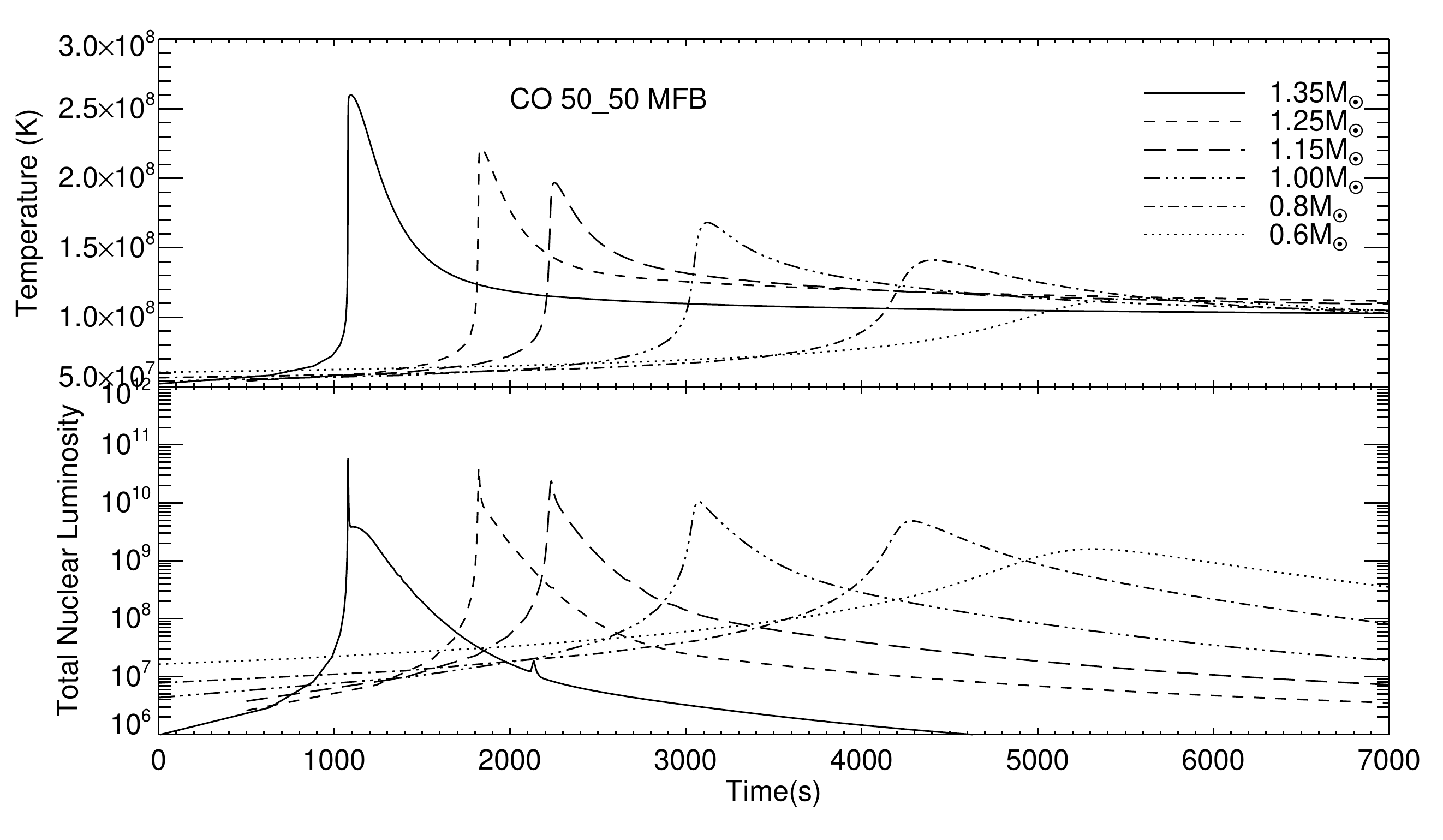}
\caption{Top panel: The same plot as in Figure \ref{MFB2575multi} but for the simulations with
50\% WD matter and 50\% solar matter and mixed from the beginning (MFB). 
While there is significantly more $^{12}$C in these simulations than in the 
25\%\ - 75\% simulations, the increased energy production, once the CNO reactions have become important,  results in less accreted mass 
and a smaller peak temperature.  Again, we have shifted the curves in time to improve their visibility. 
Bottom panel: the same plot as shown in the bottom of  Figure \ref{MFB2575multi} but for the simulations
with 50\% WD matter and 50\% solar. The small ``glitch'' in the 1.35M$_\odot$ sequence
at a time of $\sim$2100 s is caused by a change in the spatial distribution of the region where nuclear burning is
occurring.}
\label{MFB5050multi}
\end{figure}

This can be seen in the next set of rows 
which give the evolutionary results for the first mixture which is 25\% WD matter and 75\% solar matter \citep{lodders_2003_aa}. 
(The composition is noted in ``bold-face'').   The rows are 
the accretion time to the beginning of the TNR,  $\tau$$_{\rm acc}$,  
and M$_{\rm acc}$ is the total accreted mass.  
The next set of rows tabulate, as a function of WD mass, the
peak temperature in the simulation (T$_{\rm peak}$) with the scaling factor in parentheses for all rows,
the peak rate of energy generation ($\epsilon_{\rm nuc-peak}$), the peak surface luminosity
in units of the solar luminosity, (L$_{\rm peak}$/L$_\odot$), the peak effective temperature (T$_{\rm eff-peak}$), the amount of
mass ejected in solar masses (M$_{\rm ej}$),  the amount of $^7$Li ejected with respect to the solar value where we have assumed that
all the $^7$Be produced in the TNR will decay to $^7$Li, (N($^7$Li/H)$_{\rm ej}$/N($^7$Li/H)$_{\odot}$),
the ratio of the ejected mass to accreted mass in percent (M$_{\rm ej}$/M$_{\rm acc}$),
and the velocity of the surface zone which is the maximum velocity in each simulation (V$_{\rm max}$).
We express the $^7$Li results in the same ratio as given by \citet[][Table 1]{hernanz_1996_aa} so as to provide a direct
comparison (see Section \ref{lithium}).  Therefore, we use the \citet{anders_1989_aa} value for N($^7$Li/H)$_{\odot}$ of $2.04 \times 10^{-9}$
although in the simulations we use the \citet{lodders_2003_aa} abundance. The specific number does not matter in the 
simulations because all of the $^7$Li is destroyed by the TNR.
In Table \ref{comparison} (Section \ref{lithium}) we compare our $^7$Li and ejecta mass predictions with those of 
\citet{hernanz_1996_aa, jose_1998_aa}, and \citet{rukeya_2017_aa} who also mixed from the beginning.  
In Section \ref{lithium} we also discuss the differences and agreements between our 3 studies.

In the following rows we tabulate exactly the same information but for the MFB simulations with 50\% WD and 50\% solar matter.
Because of the increase in initial $^{12}$C abundance, once the accreting material gets sufficiently hot for CNO burning rather than
the initial $p-p$ chain, which now includes the $pep$ reaction: $p + e^{-}
+p \rightarrow d + \nu$ as discussed in \citet{starrfield_2009_aa}, the increased energy generation per unit accreted mass reduces the time to the TNR and the amount
of accreted mass.  Interestingly, the peak temperature during the TNR is roughly the same for both mixtures.  However, the peak rate of energy generation
is considerably higher in the 50\% WD - 50\% solar mixture because of the increased $^{12}$C abundance.  The remaining evolutionary parameters are also higher
for the 50\% WD - 50\% solar mixture.  The most material ejected at the highest velocities occurs for the 50\% WD - 50\% solar  simulation on the
1.35 M$_\odot$ WD.  However, the amount of ejected mass, $4.0 \times 10^{-7}$M$_\odot$ is far lower than the typical ejecta mass
estimates for CNe and so are the associated ejecta velocities \citep{warner_1995_aa, gehrz_1998_aa, bode_evans_08, starrfield_2012_basi}.

In the first three figures, we plot the evolutionary results for the MFB simulations.  The top panel of Figure \ref{MFB2575multi} shows the variation of temperature with time
for the zone where peak conditions occur for all 6 CO WD masses.  Our composition for these simulations is
25\% WD - 75\% solar matter, identified on the plot as 25\_75 MFB.  The WD mass is identified in the legend on the figure.  We use the same 
line identifiers for WD mass in all the plots in this paper.  As expected, the most massive WD reaches the highest peak temperature.  
 We have offset each evolutionary sequence in time so as to clearly show the rise to maximum temperature and decline.  
The time axis is chosen to emphasize the major features in the evolution of each of the WD simulations. 
Peak temperature is reached a few hundred seconds after the increasing temperature exceeds $10^{8}$K.  The rise in
temperature ends when virtually all the light nuclei in the convective region have become 
positron-decay nuclei ($^{13}$N, $^{14}$O, $^{15}$O, and $^{17}$F) and no further proton captures can occur on $^{14}$O and $^{15}$O
until they have decayed \citep[][]{starrfield_1972_aa, starrfield_2016_aa}. The simulation for the
0.6 M$_\odot$ WD shows that the temperature has just reached the peak after 8000 s of evolution.
We follow each of the simulations through peak temperature and  its decline to values where no further nuclear burning
is occuring in the outer layers.

Figure \ref{MFB5050multi} shows the same two plots as in Figure \ref{MFB2575multi} but for the composition with 50\% WD matter and
50\% solar matter (50\_50 MFB in the plots).  Note the difference in the time axes between Figure \ref{MFB2575multi} and Figure \ref{MFB5050multi}.  As seen for both compositions, not only is the 
peak temperature an increasing function of WD mass, the rise and decay times are also  functions of WD mass.  For example in
Figure \ref{MFB5050multi}, the rise time for the 1.35 M$_\odot$ WD is tens of seconds while that for the 0.6 M$_\odot$ WD is
more than 4000 s.  

The bottom panels of both Figure \ref{MFB2575multi} and Figure \ref{MFB5050multi} show the evolution of the
total nuclear luminosity (in units of the solar luminosity) as a function of time for each composition.  Again, the rise time for the most massive WDs is
much shorter than for the lower mass WDs.  Clearly, however, for these two compositions, the peak nuclear energy generation is
nearly the same for the most massive WDs.   The nuclear energy in the 50\% WD - 50\% solar simulations declines faster 
than in the 25\% WD - 75\% solar simulations because the
ejection velocities are larger and the temperatures are dropping more rapidly.  The sharp spike is characteristic of all our enriched carbon simulations.  There is a steep rise
to maximum nuclear luminosity as the expanding convective region encompasses  more of the accreted layers,
carrying the  $\beta^+$- unstable nuclei to the surface.  In addition, most of the CNO nuclei in the envelope are now
$\beta^+$- unstable nuclei and any further rise in nuclear luminosity depends on these nuclei decaying.  Their
decay at the surface causes the peak energy generation in the surface mass zones to
exceed $10^{14}$ erg gm$^{-1}$s$^{-1}$ and results in an immediate expansion of the WD outer layers.

\begin{figure}[htb!]
\includegraphics[width=1.0\textwidth]{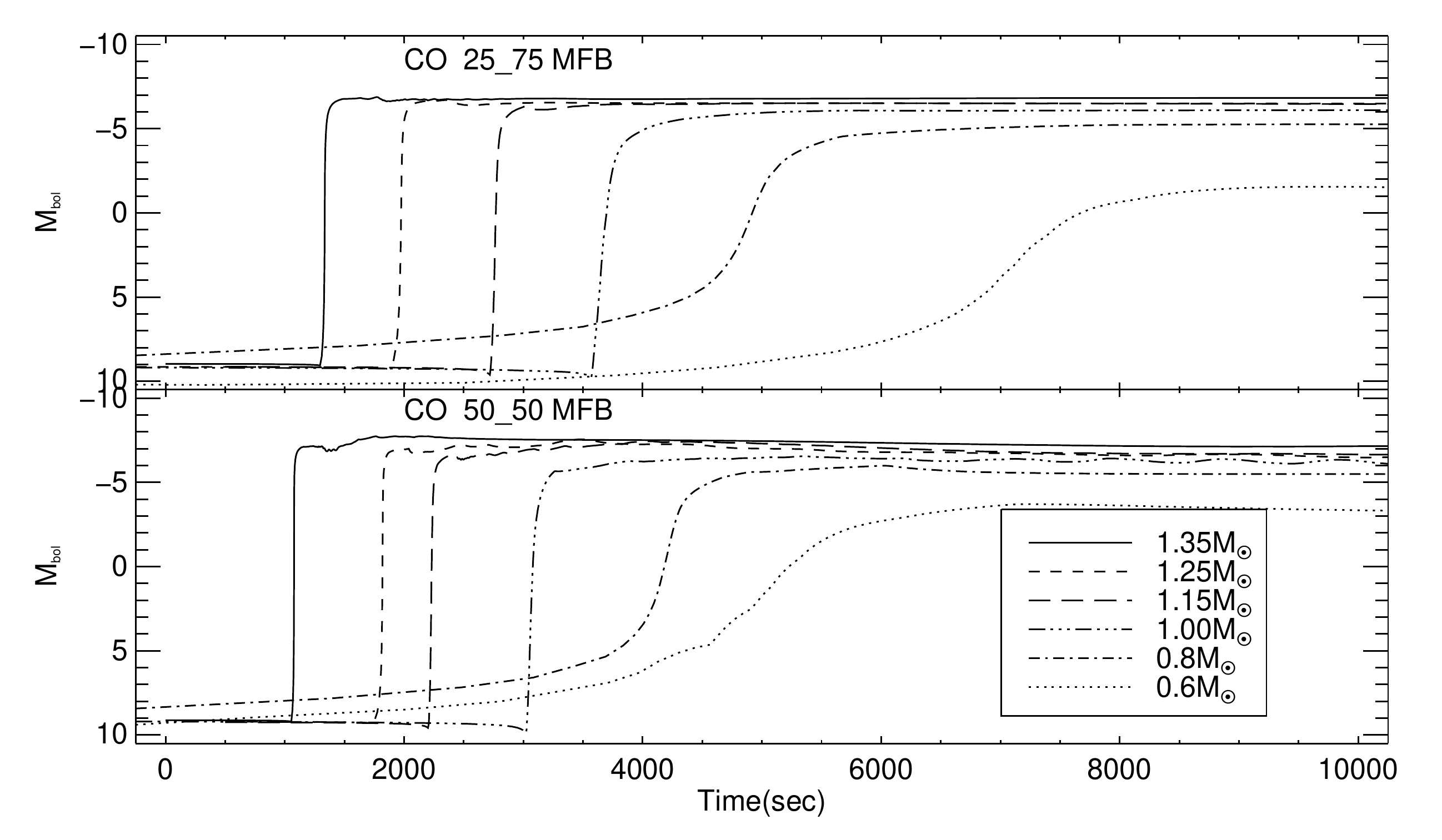}
\caption{Top panel: the variation with time of the absolute bolometric magnitude for
the simulations where we mixed a composition of 25\% WD matter and
75\% solar from the beginning (MFB).  While the simulations
on the more massive CO WDs reach values that agree with observations, those on the
lower mass WDs are too faint to agree with the observations.  Bottom panel: The same plot as in the top panel but for a mixed
composition of 50\% WD matter and 50\% solar matter.  Again the simulations on
lower mass CO WDs do not reach to values that agree with typical CN observations where M$_{\rm {Bol}}$ is around -7 or higher.
The small variations are real and suggest oscillatory behavior in the light curves but these times are normally before
the nova is discovered. }
\label{MFBlightcurve}
\end{figure}

In Figure \ref {MFBlightcurve}  we show the initial evolution of the 
bolometric magnitude as a function of time for both mixtures.  The rapid rise to maximum is caused by the intense heat from the
decays of the $\beta^+$- unstable nuclei that have reached the surface on the convective turn-over time ($\sim$ 200 s).  In contrast, the absolute visual magnitudes
for these simulations climb slowly in time as the expanding surface layers cool to $\sim 10^4$K. (They are not shown in order to prevent clutter in the figures.)  
The outermost zones reach this temperature when the surface radii have expanded to about $10^{12}$ cm and we end the evolution.  At this time the outermost
layers have become optically thin and, if they have reached escape velocity, are expanding ballistically.   We do not follow the simulations longer because
the density in these layers has dropped below $\sim 10^{-12}$ gm cm$^{-3}$.  

We note that attempts to predict the evolution of the light curve at later times typically 
use the Rosseland Mean which is a transparency mean (1/opacity) combined with a black-body source function.  
However, the atmospheres of CN after maximum do not resemble black-bodies.

We end this section by emphasizing that a key parameter affecting the evolution is the initial $^{12}$C abundance in the accreted material \citep{hernanz_1996_aa}.  This nucleus
is a catalyst in the CNO cycle,  and the MFB prescription implies a much higher initial $^{12}$C concentration than starting the simulation with just a solar
composition.  By increasing the amount of $^{12}$C with respect to hydrogen, once the CNO cycle becomes important, the rate of energy production is increased and the 
temperature in the nuclear burning region increases rapidly, per unit accreted mass, to the peak of the TNR.  Thus, less mass is accreted
than if the mixture  had a solar composition (this is shown in Section \ref{solar}). Reducing the amount of accreted mass at the time
of peak temperature in the nuclear burning region, results in a lower density and temperature, and, thereby, less degeneracy.  If the material is less
degenerate, then it expands earlier in the TNR and, in combination with the formation of the $\beta^+$-unstable nuclei halts the rising temperature.  
Since the peak temperature is lower, the peak rate of energy generation is lower causing 
the total amount of energy produced during the TNR to be smaller.  In consequence, too little mass is ejected,
at too low velocities, and the properties of the simulations do not resemble typical CNe or RNe observations \citep{warner_1995_aa, bode_evans_08, starrfield_2012_basi}.

\section{Simulations with the composition mixed during the Thermonuclear Runaway}
\label{MDTNR}

As already emphasized, the treatment of the composition of the accreted material has changed in this study compared to our prior work.  
In our last study, we assumed that the mixing of WD and accreted material occurred from
the beginning of the simulation and only used a composition of 50\% WD (ONe) and 50\% solar material \citep{starrfield_2009_aa}. 
We began this study using this procedure but assumed a CO composition and found as reported both in Section \ref{MFB} and previously in \citet[][for a CO composition]{starrfield_1997_aa}, that the results
(ejected mass and ejecta velocities) were to small to agree with the observations.  

In order to increase the amount of accreted material, therefore, we now
use the results of multi-dimensional simulations as guides.  These studies show that sufficient material is dredged-up into the accreted envelope from the outer layers of the WD by
convectively associated instabilities when the TNR is well underway \citep{casanova_2010_ab, casanova_2011_ab, casanova_2016_ab, casanova_2018_aa, jose_2014_aa}.  We simulate their calculations by first accreting 
a solar mixture \citep{lodders_2003_aa} until the temperature in the nuclear burning region exceeds
$\sim 7.0 \times 10^7$K and $\sim$ 96\% of the accreted material isconvective.  
At this time, we switch the composition of the accreted layers to a mixed composition (both abundances and the associated
equations of state and opacities) and subsequently evolve the simulation through peak temperature
and decline.  It typically takes NOVA less than 2 s of ``star'' time (but many time steps) to adjust to the new composition.
A similar technique has already been used by \citet{jose_2007_aa} in their study of the ``First Nova Explosions.'' They
explored a variety of time scales for mixing the WD material into the accreted layers, once convection was
underway, and found that using short time scales was warranted.   We choose an ``instantaneous'' time for mixing
both because it is easily reproducible and because it is not in disagreement with their results.

\subsection{Solar Accretion}
\label{solar}

\begin{figure}[htb!]
\includegraphics[width=1.0\textwidth]{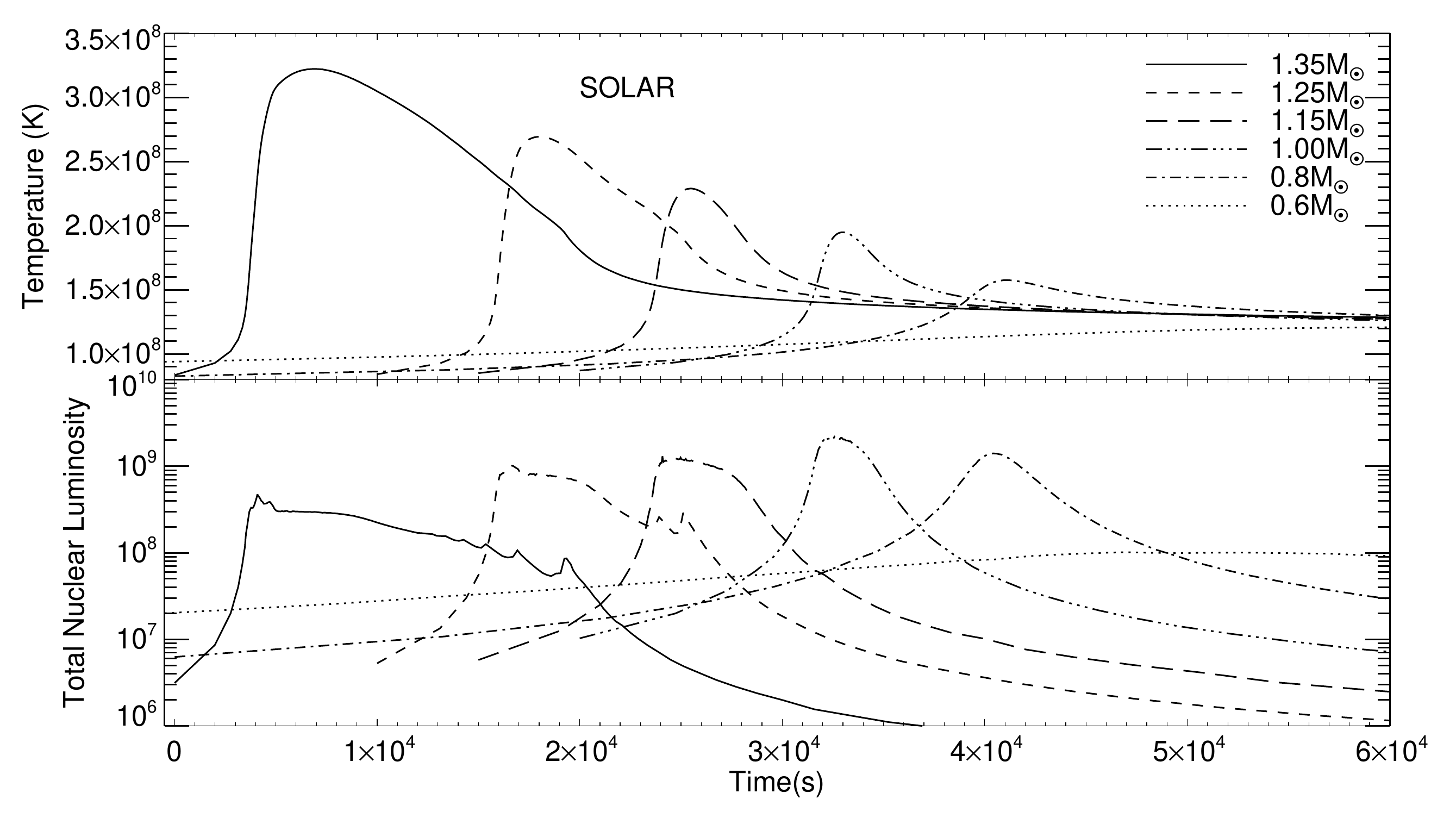}
\caption{Top Panel: the same plot as in Figure \ref{MFB2575multi} but for the simulations with
accretion of only solar material.  These simulations are used to determine the amount
of mass accreted before switching to a mixed composition.  Nevertheless, we follow
them through the explosion. The time axis is much longer than for any of the other 
simulations because of the slow evolution of these sequences.  Note that the 0.6 M$_\odot$ simulation has not yet
reached peak temperature after $5 \times 10^4$s of evolution although it started from the same beginning
temperature as the simulations for other CO WD masses.  Bottom panel: the same plot as shown in the bottom panel
of Figure \ref{MFB2575multi} but for the simulations
that accrete only solar matter and are assumed not to have mixed with WD matter.  The
``glitches'' seen in the more massive WD evolution are caused by the convective region
changing its spatial distribution, with respect to the mass zones,  as the material expands. Unlike all the other simulations,
the total energy increases as the WD mass decreases. This is because the accreted mass
has declined with increasing WD mass and there is less material involved in the evolution. In contrast,
both peak temperature and peak energy generation do increase with increasing CO WD mass as is
shown in Table \ref{evolCOMDTNR}.}
\label{COsolarmulti}
\end{figure}

In this subsection, we present the evolution of just the solar accretion phase of the study.  
We then follow that with subsections describing the simulations assuming the mixed compositions.
The initial conditions and evolutionary results are presented in Table \ref{evolCOMDTNR}.  The variables in the tables are the same as
already described for Table \ref{evolCOMFB}.  The initial conditions for each of the 6 CO WD masses are given in the first 3 rows.  The values in
these rows are identical to the first 3 rows in Table  \ref{evolCOMFB} and are repeated here only for consistency.
The next two rows give the accretion time to the beginning of the TNR,  $\tau$$_{\rm acc}$,  
and M$_{\rm acc}$ is the total accreted mass at that time.   These values are those used both for the solar accretion simulations and, later, for the
two mixed composition simulations for each of the listed WD masses (Section \ref{MDTNR}).  We begin each of the sets of simulations with the composition listed in ``bold-face''. 
As in the MFB simulations, we use 150 mass zones, with the mass of the zone decreasing outward in radius, and accrete at $1.6 \times 10^{-10}$M$_\odot$ yr$^{-1}$.  

We can see the immediate effects of accreting a solar composition instead of a mixed composition.
Comparing the results given in Table \ref{evolCOMDTNR} to those in Table \ref{evolCOMFB}, the reduced amount of $^{12}$C in the solar accretion
simulations significantly increases the amount of accreted mass.  For example, comparing the solar accretion simulation to the 25\% WD  -75\% solar (MFB) simulation, we find that
about twice as much mass is accreted at 0.6 M$_\odot$ and a factor of 3 times more mass at 1.35 M$_\odot$.  Peak temperature is higher for
all WD masses in the Solar accretion simulation as compared to the MFB simulations.  The increased mass and degeneracy at the bottom of the accreted material clearly compensates for the larger amount of $^{12}$C in the MFB simulations.

Figure \ref{COsolarmulti} (top panel) shows the evolution of the temperature with time for the zones where peak conditions in the TNR occur for all the CO WD masses 
accreting just a solar composition.  Although there is more accreted mass in each of the simulations, the temperature evolution is extremely slow as shown by the time axis.  
While the 1.35M$_\odot$ simulation takes  $\sim 2 \times10^4$ s to evolve through the peak and decline of the TNR,  the 0.6M$_\odot$ simulation is still on
the rise after $\sim 5 \times 10^4$ s.  In contrast, the equivalent MFB simulations take a far shorter time (shown in Figure \ref{MFB2575multi} and Figure \ref{MFB5050multi}) to evolve through the peak as do the simulations to be reported on in the next subsection.

Figure \ref{COsolarmulti} (bottom panel) shows the variation in 
total nuclear energy generation around the peak of the TNR.  It should be compared with the bottom panels in Figures \ref{MFB2575multi} and \ref{MFB5050multi}.
The rise to peak nuclear energy generation is extremely slow and the decline
is also slow.  In addition, the peak is more than a factor of 10 lower than in the MFB simulations for the same WD mass. 
The ``glitches'' seen in the more massive WD evolution are caused by the convective region
changing its spatial distribution, with respect to the mass zones, as the material expands. Unlike all the other simulations,
the total energy as a function of time increases as the WD mass decreases. This is because the total accreted mass
has declined with increasing WD mass and there is less material involved in the evolution. However,
both peak temperature and peak energy generation do increase with increasing CO WD mass as is
shown in Table \ref{evolCOMDTNR}.

\begin{figure}[htb!]
\includegraphics[width=1.0\textwidth]{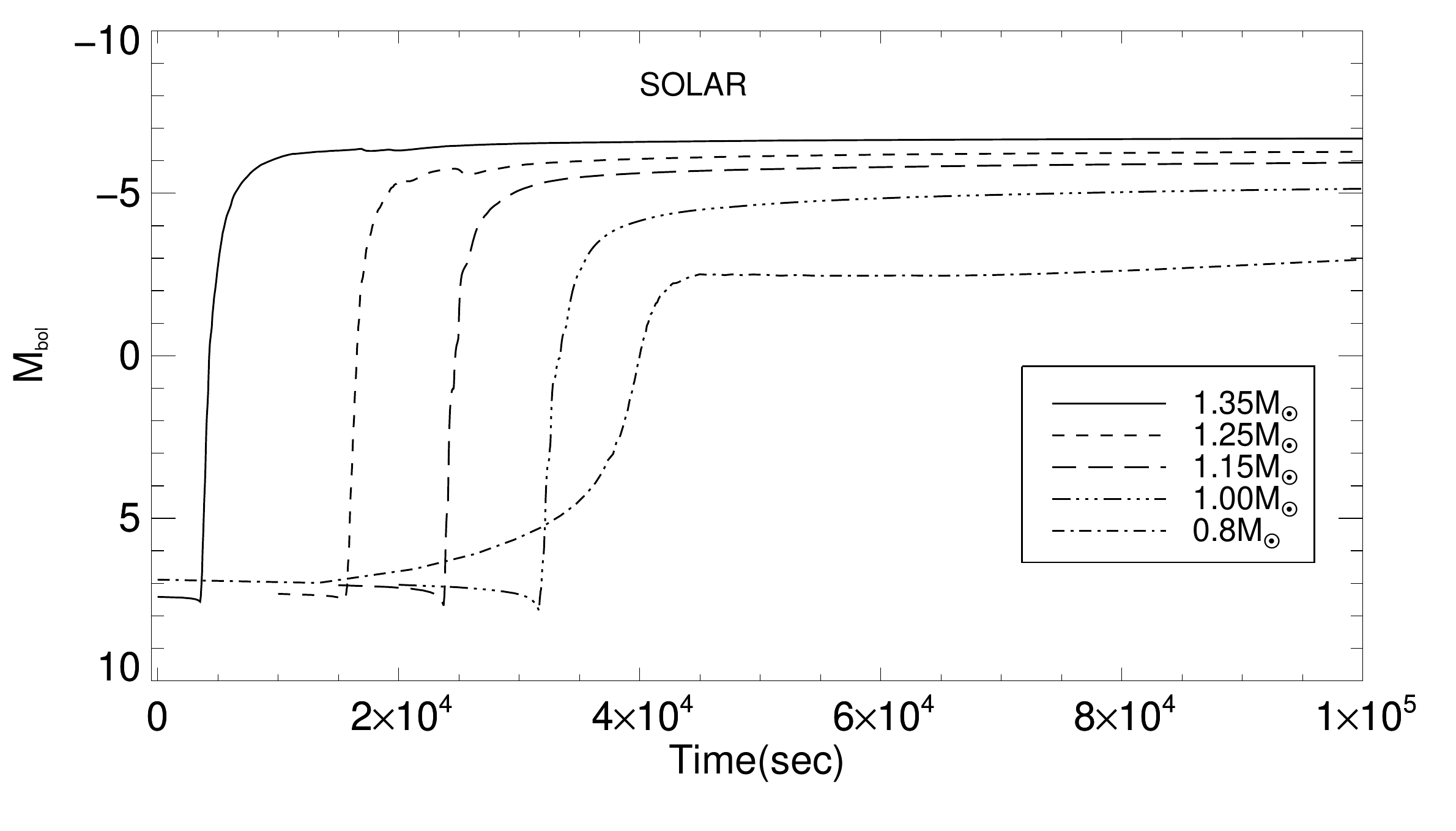}
\caption{The same plot as shown in Figure \ref{MFBlightcurve} but for the solar composition
simulations.  Note that only the simulations on the most massive CO WDs reach peak values close to
those that are observed.  They also evolve extremely slowly  compared to the mixed compositions (both MFB and MDTNR) as can
be seen on the time axis.  We do not plot the evolution of the 0.6 M$_\odot$ simulation since it is still rising after $10^5$s. }
\label{COsolarmbolt}
\end{figure}

Figure \ref{COsolarmbolt} is the solar accretion analog of Figure \ref {MFBlightcurve},
showing the time evolution of the bolometric magnitude for the solar accretion simulations.  Peak M$_{\tt{bol}}$ is an increasing function of WD mass but even
the simulation on the most massive WD does not reach values that are observed in a typical CN outburst of $\sim$ -8.  These, however,  might fit 
some of the slowest and faintest CNe shown in \citet{kasliwal_2011_aa}.  

Alternatively, mixing may occur in these CNe but, if the outer layers of the WD
consist of material that has undergone previous CN outbursts, and the outburst has left a helium enriched layer behind,  it will
be helium enriched material that is mixed into the accreted layers in addition to $^{12}$C
enriched material. In fact, spectroscopic studies of CNe ejecta show that this material is strongly enriched in helium
and to amounts that suggest that helium has been mixed up from below and is not just the
residue of the hot-hydrogen burning reactions that drove the TNR. 

Finally, we note that the consequences of accretion of solar material onto WDs with a larger variation in mass and 
mass accretion rates, and where {\bf no} mixing of WD with accreted material was assumed, has  been published elsewhere \citep{starrfield_2012_basi, starrfield_2012_aa,
newsham_2014_aa}.  They found that a TNR occurred for all WD masses and mass accretion rates.

\subsection{The Simulations using Compositions Mixed During the TNR}
\label{MDTNR}

\begin{deluxetable}{@{}lcccccc}
\tablecaption{Initial Parameters and Evolutionary Results for Accretion onto CO WDs: 
Mixing During the Thermonuclear Runaway (MDTNR)  
\label{evolCOMDTNR}}
\tablewidth{0pt}
\tablecolumns{7}
\tablehead{ \colhead{CO WD Mass (M$_\odot$): }&
\colhead{0.6}&
\colhead{0.8} &
\colhead{1.0} &
\colhead{1.15} &
\colhead{1.25} &
\colhead{1.35}} 

\startdata
Initial:  L/L$_\odot$($10^{-3}$)&4.8&4.8&4.9&4.7&4.8&5.5\\
Initial: R($10^3$km)&8.5&6.8&5.3&4.2&3.4&2.3\\
Initial: T$_{\rm eff}$($10^4$K)&1.4&1.6&1.8&2.0&2.2&2.7\\
\hline
 $\tau$$_{\rm acc}$($10^5$ yr)&19.8&9.9&5.1&2.4&1.6&0.6\\
M$_{\rm acc}$($10^{-5}$M$_{\odot}$)&31.4&16.0&8.1&3.9&2.6&1.0\\
\hline
{\bf Solar mixture}&&&&&&\\
\hline
T$_{\rm peak}$($10^8$K)&1.2&1.6&2.0&2.3&2.6&3.0\\
$\epsilon_{\rm nuc-peak}$($10^{14}$erg gm$^{-1}$s$^{-1}$)&$0.032$&$0.73$&$1.6$&$1.8$&$1.6$&$1.8$\\
L$_{\rm peak}$/L$_\odot$ ($10^4$)&3.9&3.8&2.1&2.7&3.2&3.9\\
T$_{\rm eff-peak}$($10^5$K)&0.9&1.6&2.0&3.4&5.3&7.7 \\
M$_{\rm ej}$($10^{-7}$M$_{\odot}$)&$6.8$&$12.0$&$0.034$&$0.33$&0.0&$0.30$\\
N($^7$Li/H)$_{\rm ej}$/N($^7$Li/H)$_{\odot}$ &$1.0\times10^{-5}$&$1.4\times10^{-5}$&$1.5\times10^{-2}$&$4.7\times10^{-3}$&$4.6\times10^{-4}$&$6.2\times10^{-3}$\\
M$_{\rm ej}$/M$_{\rm acc}$(\%)&0.2&0.8&$\sim$0.0&$\sim$0.0&0.0&0.3\\
V$_{\rm max}$($10^2$km s$^{-1}$)&3.6&4.5&3.8&5.1&0.0&4.8\\
\hline
{\bf 25\% White Dwarf - 75\% Solar}&&&&&&\\
\hline
T$_{\rm peak}$($10^8$K)&1.3&1.7&2.0&2.4&2.8&3.4\\
$\epsilon_{\rm nuc-peak}$($10^{16}$erg gm$^{-1}$s$^{-1}$)&$0.66$&$0.65$&$1.2$&$3.4$&$11.4$&$34.3$\\
L$_{\rm peak}$/L$_\odot$ ($10^5$)&0.7&8.5&2.8&2.3&3.3&7.3\\
T$_{\rm eff-peak}$($10^5$K)&1.1&2.4&3.2&8.4&8.0&10.5 \\
M$_{\rm ej}$($10^{-6}$M$_{\odot}$)&$0.49$&$2.9$&$4.3$&$12.8$&$20.8$&$4.6$\\
N($^7$Li/H)$_{\rm ej}$/N($^7$Li/H)$_{\odot}$ &$78.0$&$5.0\times10^2$&$1.6\times10^{3}$&$2.8\times10^{3}$&$3.0\times10^{3}$&$3.5\times10^{3}$\\
M$_{\rm ej}$/M$_{\rm acc}$(\%)&0.2&2&5&33&80&46\\
V$_{\rm max}$($10^3$km s$^{-1}$)&0.4&1.4&2.8&2.8&3.4&1.4\\
\hline
{\bf 50\% White Dwarf - 50\% Solar}&&&&&&\\
\hline
T$_{\rm peak}$($10^8$K)&1.4&1.8&2.2&2.5&3.4&4.3\\
$\epsilon_{\rm nuc-peak}$($10^{17}$erg gm$^{-1}$s$^{-1}$)&$0.012$&$0.085$&$0.47$&$1.7$&$26.0$&$298.0$\\
L$_{\rm peak}$/L$_\odot$ ($10^6$)&3.0&2.6&3.4&2.0&2.7&3.9\\
T$_{\rm eff-peak}$($10^5$K)&1.5&3.2&4.2&8.3&8.3&43.0 \\
M$_{\rm ej}$($10^{-5}$M$_{\odot}$)&$16.0$&$11.0$&$6.3$&$3.4$&$2.3$&$0.86$\\
N($^7$Li/H)$_{\rm ej}$/N($^7$Li/H)$_{\odot}$ &$1.9\times10^{2}$&$7.4\times10^{2}$&$1.6\times10^{3}$&$2.8\times10^{3}$&$3.6\times10^{3}$&$3.4\times10^{3}$\\
M$_{\rm ej}$/M$_{\rm acc}$(\%)&51&69&78&87&90&88\\
V$_{\rm max}$($10^3$km s$^{-1}$)&2.5&4.3&6.4&6.6&6.5&5.6\\
\enddata

\end{deluxetable}

We now take the results for each CO WD mass from the evolution reported in the last subsection and
switch to a mixed composition when the peak temperature in the simulation has reached 
$\sim 7.0 \times 10^7$ K and convection is well underway but has not yet reached the surface.  
Once we have switched the composition, we continue the evolution, without assuming any further accretion, through peak temperature of the TNR
and the following decline in temperature to where there is no further nuclear burning in the
outer layers.  We use the same two mixtures (either 25\% WD and 75\% solar material or 50\% WD and 50\% solar material) 
as used in the MFB simulations.   These two sets of simulations are identified on the plots as either 25\_75 Mixing During the TNR (25\_75 MDTNR)
or ~50\_50 Mixing During the TNR (50\_50 MDTNR).

We emphasize that the same {\it initial} model (thermal structure, spatial structure, and amount of accreted mass distributed through the
same number of mass zones) is used for the two sets of
simulations at the time we switch to a mixed composition.  
The violence of the resulting evolution now depends on the amount of $^{12}$C in the accreted layers (solar plus WD) {\it after} the switch in composition.  
Since, for the same temperature and density,  increasing the $^{12}$C abundance increases the rate of energy generation, 
the simulations evolve much more rapidly, and reach higher peak values than the equivalent MFB simulation (for the same WD mass and mixed composition).  
This result is in contrast to the previously described MFB accretion phases (Section \ref{MFB}) where a higher initial $^{12}$C abundance resulted in a weaker explosion.  
Peak temperature is reached about one hundred seconds after the temperature exceeds $10^{8}$K.  The rise in
temperature ends because virtually all the light nuclei in the convective region have become 
positron decay nuclei.

We follow each of the simulations through peak nuclear burning, peak temperature, and decline.  We end the simulation when the outer layers
have reached radii of $\simeq$ few $\times 10^{12}$ cm.  At this radius they have started to become optically thin and,
in some of the simulations, the material has reached escape  velocity and the density has declined to below $\sim 10^{-12}$
gm cm$^{-3}$.

\begin{figure}[htb!]
\includegraphics[width=1.0\textwidth]{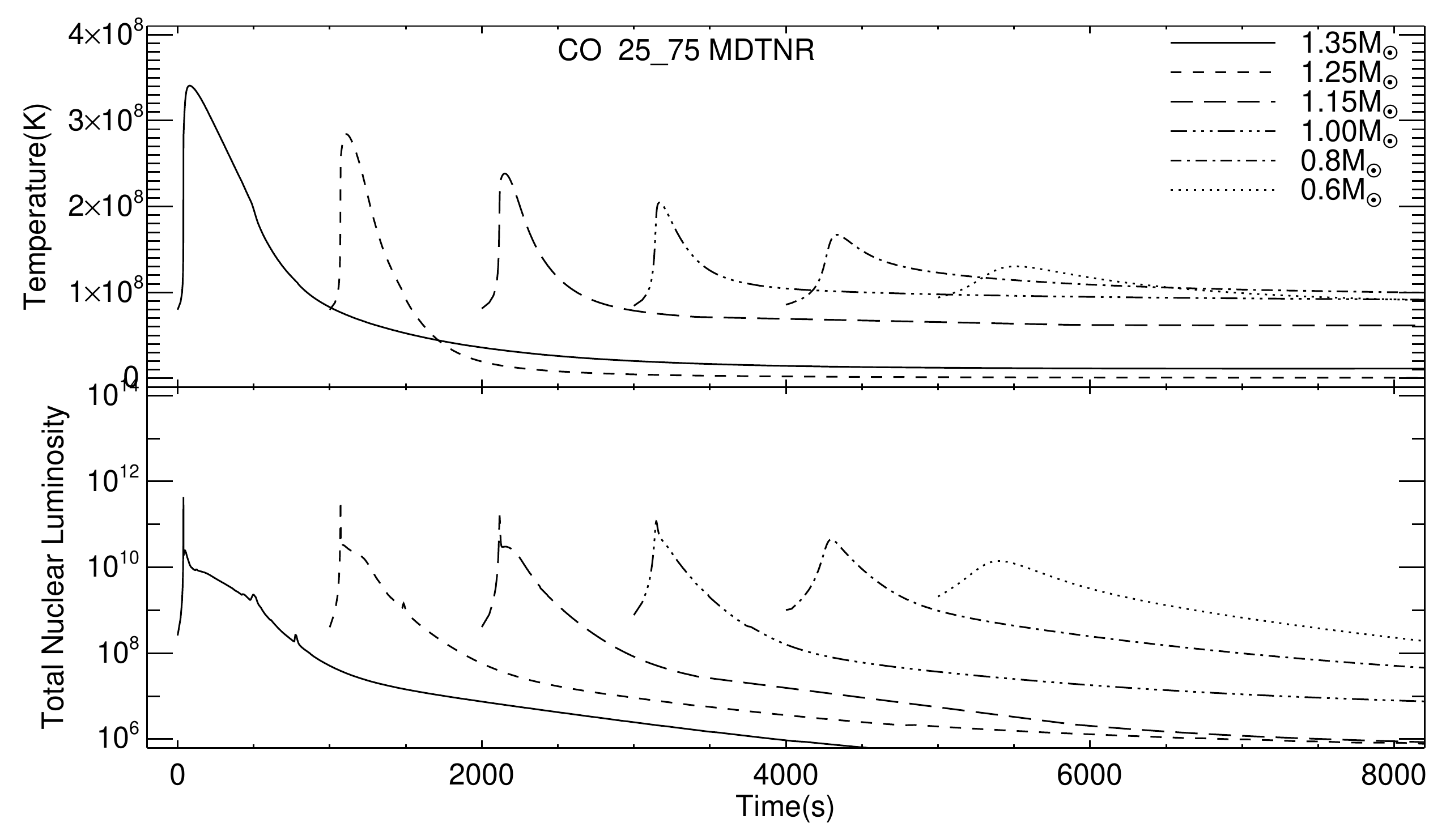}
\caption{Top panel: the variation with time of the temperature in those mass zones near the interface
between the outer layers of the WD and the accreted plus WD matter 
for the simulations with 25\% WD material and 75\% solar.   This plot is the analog of the top panel of Figure \ref{MFB2575multi}.
The results for all six simulations are shown (the WD mass is identified in
the inset).  The curve for each sequence has
been shifted in time to improve its visibility. As expected, the peak temperature
achieved in each simulation is an increasing function of CO WD mass. 
Bottom panel: same as the bottom panel of Figure \ref{MFB2575multi} but for the simulation with 25\% WD-75\% solar. 
The small ``glitches'' that appear on the decline are caused by
convection moving in and out and bringing in small amounts of fresh nuclei to the nuclear 
burning regime.}
\label{figuremultiaa}
\end{figure}

\begin{figure}[htb!]
\includegraphics[width=1.0\textwidth]{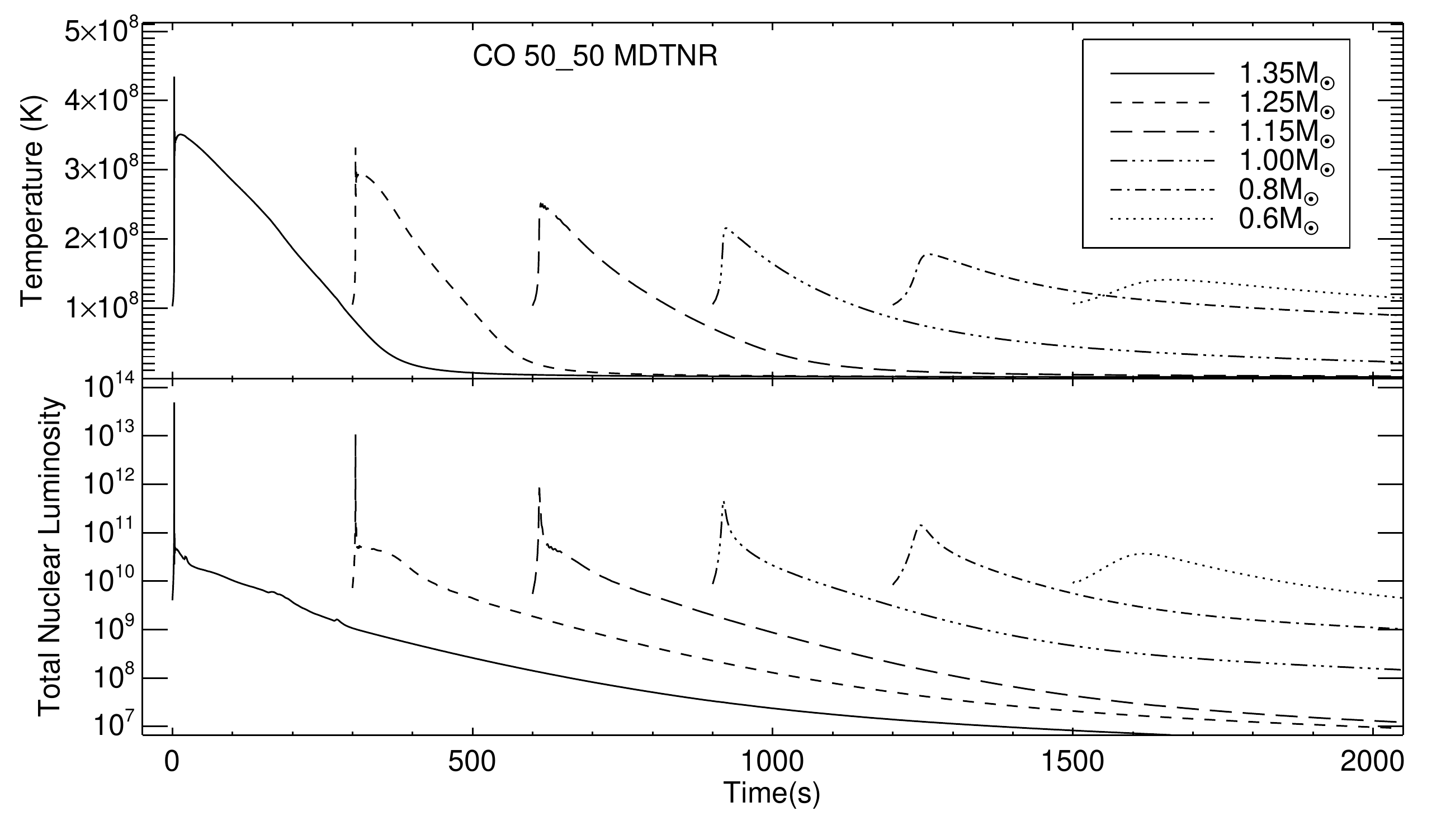}
\caption{Top panel: same as Figure \ref{figuremultiaa} but for the simulations with 
50\% WD matter and 50\% accreted matter.  Because of the increased $^{12}$C, the rate of energy generation is larger for a given temperature
and density, and these simulations evolve much more rapidly than the simulations with a lower $^{12}$C
abundance.  Therefore, the evolution time for this series of sequences is significantly shorter than that in Figure \ref{figuremultiaa}. The
extremely rapid increase and decrease in temperature indicate that the simulation formed a shock wave in the
zone where peak temperature occurred. Bottom panel: the same plot as in Figure \ref{figuremultiaa} but for the simulation with
50\% WD matter and 50\% accreted matter.  Note that the horizontal and vertical axes
differ in these two plots.  Because of the much larger amount of $^{12}$C in these simulations, 
the evolution is more extreme and faster than for those simulations with a smaller amount
of $^{12}$C.}
\label{figuremultiab}
\end{figure}

As discussed in the last subsection, the first set of evolutionary results in  Table \ref{evolCOMDTNR} shows the consequences of 
following the solar accretion simulation through the TNR, without any mixing, and its return to near quiescence.  
The next two sets of rows provide exactly the same information but for the mixed composition simulations with 25\% WD and 75\% solar matter, and followed
below by the results for the 50\% WD and  50\% solar matter simulations.   Comparing the 25\% WD-75\% solar MDTNR simulations to the
50\% WD-50\% solar MDTNR simulations, the large enrichment of $^{12}$C in the simulations with more WD matter causes a more extreme set of evolutionary results.

The plots of temperature versus time for these two sets of simulations are given in Figures \ref{figuremultiaa} and \ref{figuremultiab}.  Both the vertical
and horizontal scales in these two figures are different.  
As noted before, the peak temperature during the TNR is an increasing function of WD mass.  The simulation with 25\% WD-75\% solar MDTNR involving a CO WD with a mass
of 1.35 M$_\odot$ reaches the highest temperature of $3.4 \times 10^8$ K, while the simulation on the lowest mass WD,
0.6 M$_\odot$, reaches the lowest peak temperature of $1.3 \times 10^8$ K (Figure \ref{figuremultiaa}).  
Comparing these values to the results for the 50\% WD-50\% solar MDTNR simulations,
we find that there is hardly any difference in peak temperature and energy generation for the lower mass WDs but the values for the 1.25 M$_\odot$ and
1.35 M$_\odot$ simulations are far larger for the more carbon enriched simulation (see Table \ref{evolCOMDTNR} and Figure \ref{figuremultiab}).  
The peak temperature for the 1.35 M$_\odot$ 50\% WD-50\% solar MDTNR simulation reaches
$4.3 \times 10^8$ K and the peak rate of energy generation is $3.0 \times 10^{19}$ erg gm$^{-1}$s$^{-1}$ in the region closest to the interface between the WD core and 
accreted plus core material.  The rise in temperature for this sequence is so rapid that a shock
forms at the interface between the accreted and WD matter and moves through the envelope in seconds.  The consequences of the shock can be seen in 
ejection velocities that exceed 5,600 km s$^{-1}$ for the most massive WDs (Table \ref{evolCOMDTNR}).  
 The sharp spike in the 1.35 M$_\odot$ 50\% WD-50\% solar MDTNR simulation shows the shock 
formation.  The luminosities and effective temperatures for the 50\% WD- 50\% solar MDTNR evolution exceed those for observed CNe
explosions. We suggest that this mixture is too extreme. 

Figures \ref{figuremultiaa} (for the 25\% WD-75\% solar MDTNR simulations) and \ref{figuremultiab} (for the 50\% WD-50\% solar MDTNR 
simulations) also show (bottom panels) the variation with time of the  total nuclear
luminosity in solar units (L$_\odot$) around the time of peak temperature.  
The glitches are caused by the spatial distribution of the convective region moving inwards and outwards and bringing in fresh partially
burned material.  Note that both the vertical and horizontal axes differ in these two plots.   Both sets of simulations show an extremely rapid rise to
maximum and a sharp decline followed by a slower decline.   The steep rise
to maximum nuclear luminosity occurs as the convective region encompasses all the accreted layers thus
carrying the  $\beta^+$- unstable nuclei to the surface and unprocessed CNO nuclei down to the nuclear burning region.

 As in the MFB evolutionary sequences, the rise time for the most massive CO WDs is
shorter than for the lower mass CO WDs. However, for these two compositions, the peak nuclear energy generation is
nearly the same for the massive WDs but decreases with decreasing WD mass for the lower mass WDs.   In contrast, as shown in
Figure \ref{figuremultiaa}  and Figure \ref{figuremultiab}, the total nuclear luminosity decreases as the CO WD mass decreases.  For all masses, the
peak is higher for the  50\% WD-50\% solar MDTNR simulations.  The extremely sharp spike for the 
1.35 M$_\odot$ 50\% WD - 50\% solar evolution  indicates that a
shock has formed. The nuclear energy in the 50\% WD - 50\% solar MDTNR simulations declines faster than in the 25\% WD - 75\% solar simulations because the
expansion velocities are larger and, therefore, the temperatures drop  more rapidly. 

Table \ref{evolCOMDTNR} shows that the peak luminosities and effective temperatures are much higher for the 50\% WD - 50\% solar MDTNR mixture and
massive WDs.  The peak luminosities for the 25\% WD - 75\% solar MDTNR simulations range from $7 \times 10^4$ L$_\odot$ for the 0.6 M$_\odot$ WD to $7.3 \times 10^5$ L$_\odot$
for the 1.35 M$_\odot$ WD, although the highest luminosity is $8.5 \times 10^5$L$_\odot$ for the 0.8 M$_\odot$ WD.  These values are not
unreasonable when compared to observations if we realize that CNe typically are discovered in outburst long after peak conditions occurred in the nuclear
burning region and we have ended the evolution. 

In contrast, the peak luminosities for the 50\% WD - 50\% solar MDTNR mixture range from $2.0 \times 10^6$L$_\odot$ to $4.0 \times 10^6$L$_\odot$ which are
too high to agree with the observations.  These high luminosities, in combination with the predicted effective temperatures which range from
from $8.3 \times 10^5$K for the 1.15 M$_\odot$ simulation to $4.3 \times 10^6$K for the 1.35 M$_\odot$
simulation, would trigger responses in some of the X-ray detectors
currently in orbit.   Such triggers have not occurred.  However, the 25\% WD-75\% solar MDTNR simulation on a 1.25 M$_\odot$ WD reaches $3.3 \times 10^5$K 
and that on a 1.35 M$_\odot$ WD reaches $2.2 \times 10^5$K which are less than seen in 
the results of some of the X-ray grating studies of CN near the peak \citep[][and references therein]{orio_2018_aa} and suggest that mixtures 
between the two that we have studied might be in better agreement with the peak luminosity predictions.

\begin{figure}[htb!]
\includegraphics[width=0.9\textwidth]{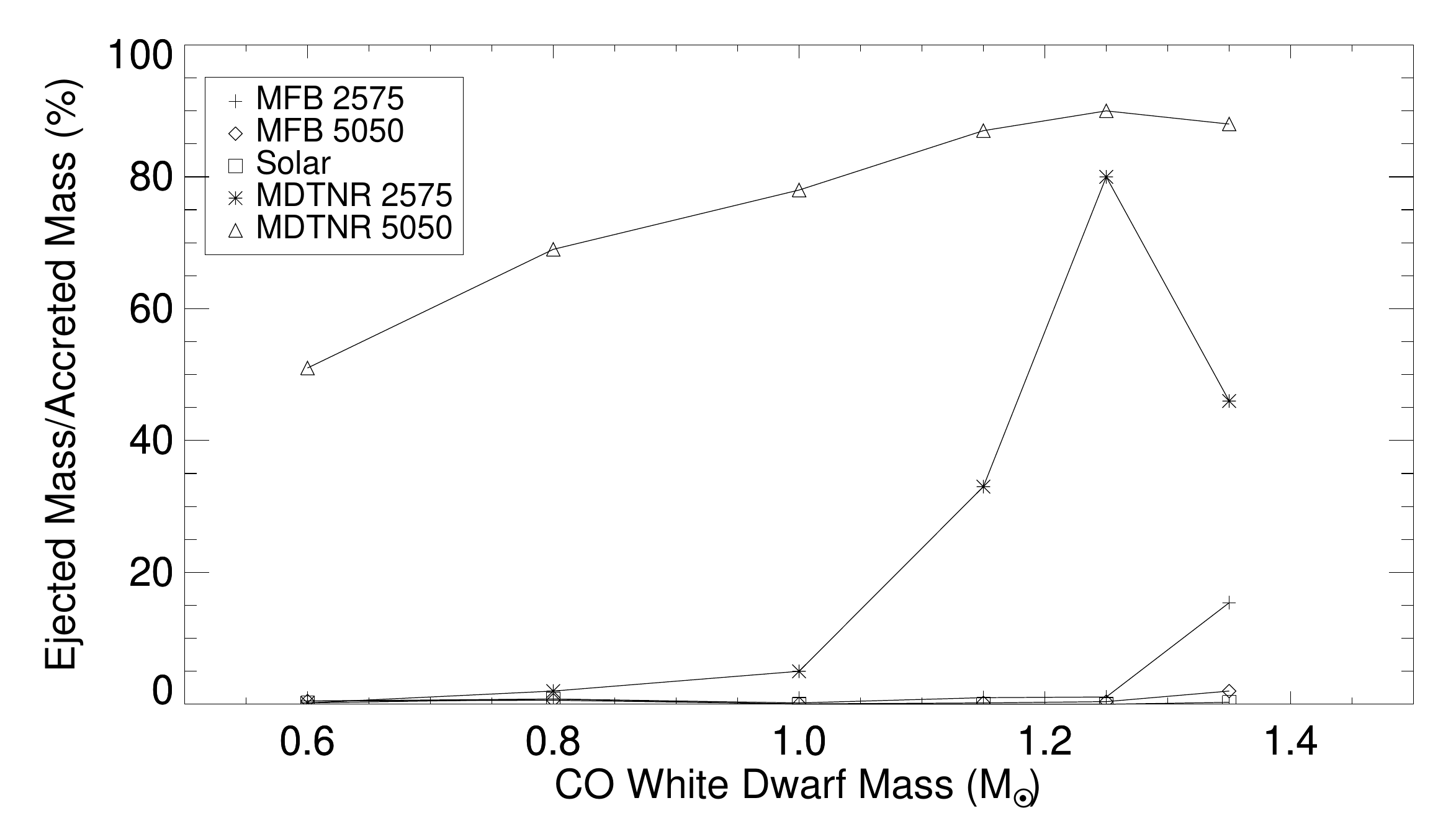}
\caption{The ratio of ejected to accreted mass (in percent) as a function of
CO WD mass.  Neither the MFB nor the solar simulations eject
much material and, thereby, the WD is growing in mass toward the
Chandrasekhar Limit.  While the 25\% WD - 75\% solar on the 1.25M$_\odot$ and
1.35M$_\odot$ simulations eject 81\% and 46\% of the accreted mass, respectively; only 25\% of the ejecta is
WD material and we predict that the WD is gaining in mass as a result of the CN outburst.  The other sequences that eject a significant amount
of material are the 50\% WD and 50\% MDTNR simulations on the massive WDs but only half of the ejecta is WD material.}
\label{figuremassej}
\end{figure}

Both Table \ref{evolCOMFB} and \ref{evolCOMDTNR} give the mass ejected by each of the simulations
along with the ratio of ejected to accreted mass (in percent) as a function of CO WD mass. 
In Figure  \ref{figuremassej}, we show the same data (the ratio of ejected to accreted mass in percent) as a function of
WD mass for all the mixed sequences that we evolved.  Clearly,  less mass is ejected
than accreted.   The only sequences that eject a significant amount
of material are the 50\% WD - 50\% solar MDTNR sequences on the most massive WDs.  
In contrast, the 25\% WD - 75\% solar sequences show
that only the 1.25 M$_\odot$ sequence ejects a sufficient amount of material so that the WD might be losing mass as a result of the TNR.  
However, for these simulations only 25\% of the ejecta is WD material.  
Nevertheless, the amount of ejected material is reduced by increasing the mass accretion rate, or the 
initial WD luminosity, or both \citep{yaron_2005_aa, starrfield_2016_aa, hillman_2015_aa, hillman_2016_aa}.  

Those sequences with 50\% WD matter and 50\% solar matter MDTNR
 could have either the WD losing mass 
(M$_{\rm{WD}}$ $\gtrsim$ 1.0 M$_{\odot}$)
or gaining mass (M$_{\rm{WD}} \lesssim$ 1.0 M$_{\odot}$) although only 50\% of the material in the accreted layers is actual WD material.  
However, the peak luminosities do not agree with the observations.   We assert, therefore, that for most
observed CO CNe less than half of the material in the accreted region comes from the WD and it is gaining 
in mass as a result of accretion, TNR, and ejection.

\begin{figure}[htb!]
\includegraphics[width=0.9\textwidth]{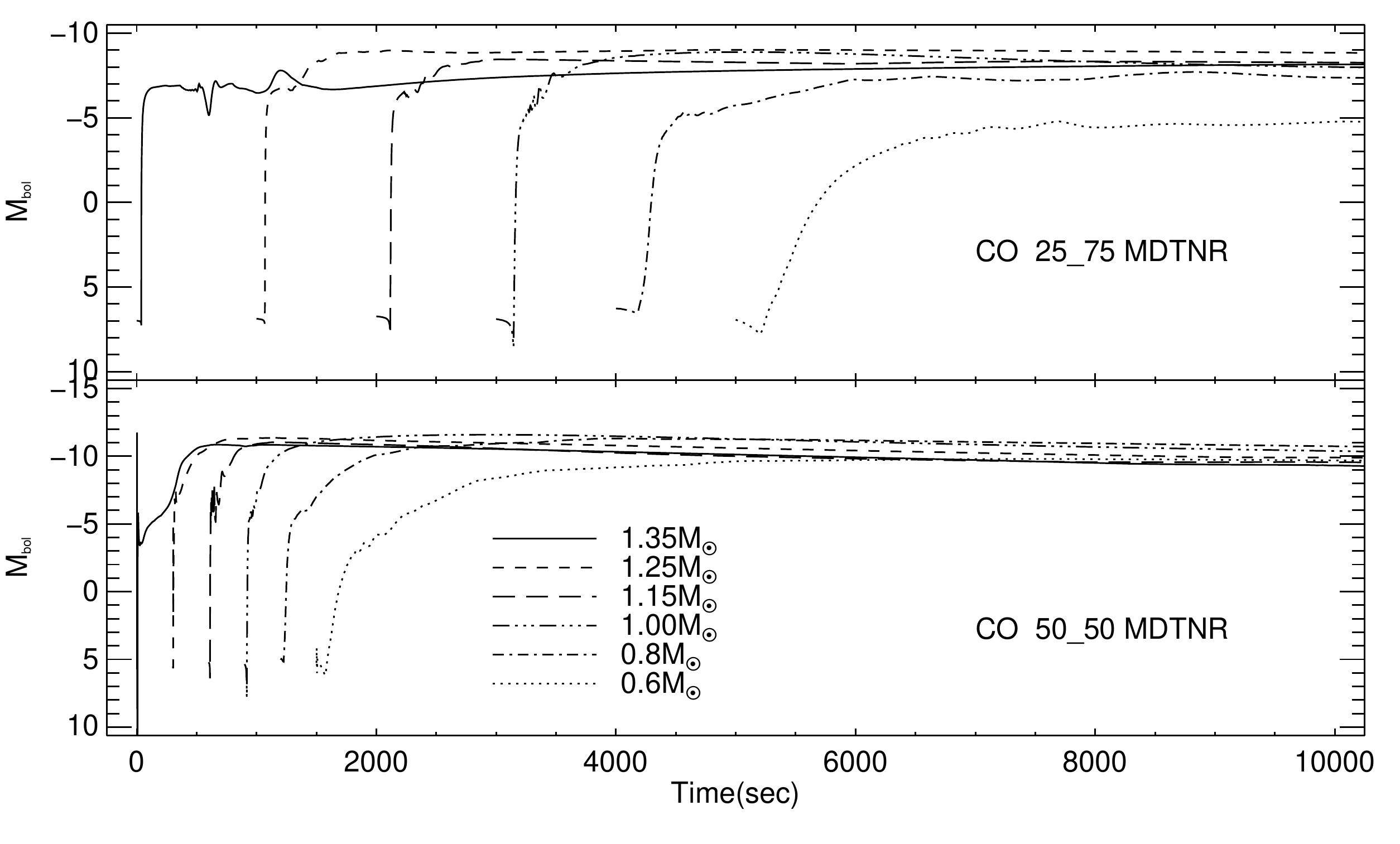}
\caption{Top panel: the variation with time of the absolute bolometric magnitude for
the simulations where we used a composition of 25\% WD matter and
75\% solar after the TNR was well underway (MDTNR).  After the initial
few hundred seconds they show a range in peak bolometric 
magnitude but there is no correlation with CO WD mass.
Bottom panel: The same plot as in the top panel but for a
composition of 50\% WD and 50\% solar.  The large amount of
$^{12}$C in these simulations drives an initial shock in the most massive
WD and the other simulations all reach a peak M$_{\rm{bol}}$
around -10 which is extremely bright for the typical CN outburst.  We,
therefore, claim that this choice of composition does not agree with observations. }
\label{MDTNRLC}
\end{figure}

We end this subsection with plots of the evolution of the bolometric magnitude (M$_ {\rm{bol}}$) with time.  Figure \ref{MDTNRLC} 
shows the first hours of the evolution of both sets of simulations (top panel: 25\_75 MDTNR and bottom panel: 50\_50 MDTNR).  
Both panels show the rapid rise in M$_ {\rm{bol}}$ as the energy produced in the nuclear burning region reaches the surface.
Subsequently,  they become roughly constant with time up to the end of the simulations.  
M$_ {\rm{bol}}$ for the 1.35 M$_\odot$ simulation with 25\% WD and 75\% solar material is lower than those of the other massive WDs
but they all appear to match observed CNe bolometric magnitudes.  
In contrast, M$_{\rm{bol}}$ for the 50\% WD - 50\% solar MDTNR simulations all lie
above those reported for typical CNe but may agree with the bright outliers seen in \citet{kasliwal_2011_aa}.  For example, their Table 5 \citep{kasliwal_2011_aa} lists
one nova in M82 with an absolute magnitude (Gunn-g) of -10.7 and one in M81 with an absolute magnitude of -9.9. Since these values refer to photometry
obtained with the Gunn-g filter and our values are absolute bolometric magnitudes, we do not attempt to put them on the same sequence.  
In addition, we end our simulations before those novae would have been discovered.  Our predicted absolute visual magnitudes rise slowly
and reach values close to those plotted for peak M$_{\rm{bol}}$ after a few hours when the T$_{\rm eff}$ has fallen below $10^4$K.

\subsection {A Detailed Look at a TNR}
\label{detailed}

In this subsection, we describe the evolution of the 25\% WD -75\% solar MDTNR simulation on a 1.35M$_\odot$ CO WD in detail.  
The gross properties of the evolution are found in Table \ref{evolCOMDTNR}.   We accrete a solar mixture until the temperature has reached a value of 
$7.5 \times  10^{7}$K, at which time the density is $9.0 \times 10^3$ gm cm$^{-3}$, the pressure is $8.8 \times 10^{19}$ dynes cm$^{-2}$, 
and the rate of energy generation is $6.2 \times 10^{15}$ erg gm$^{-1}$s$^{-1}$.  We then switch to the mixed composition
and continue through the peak of the TNR and decline in temperature.  

It takes this simulation only 0.2 s to adjust to the new composition. The convective region extends for 49 km, 
from the Core-Envelope Interface (CEI: the mass zone where pure WD material connects to the accreted plus WD zones), toward the surface.  
Because the mass of the zones decrease with increasing radius, almost 96\% of the accreted layers are in the convective region so that when
convection reaches the surface (about 30 s later) there is no major change in the composition. 
At switch-over, the mass fraction of
$^3$He is $2.6 \times 10^{-5}$, $^2$H is $2.1 \times 10^{-5}$, and $^7$Be is zero.  
We report the evolution of $^3$He because it is converted to $^7$Be through the $^3$He($\alpha$,$\gamma$)$^7$Be reaction.  
$^7$Be then decays to $^7$Li with a $\sim$53 d half-life.

After 41.4 s of evolution (times are given since the
beginning of the switch in composition), the sequence reaches a temperature of $10^8$K just above the CEI, and after 50.4 s it reaches a peak
rate of energy generation of $3.4 \times 10^{17}$ erg gm$^{-1}$s$^{-1}$ at the CEI.  
The temperature in this mass zone is $2.5 \times 10^8$K, the density is $2.5 \times 10^3$ gm cm$^{-3}$,  and convection has reached the surface
layers of the WD.  
At 50.55 s (the full printout shows) the temperature has risen to $2.7 \times 10^8$K, the density and nuclear energy generation have
have fallen to  $2.0 \times 10^3$ gm cm$^{-3}$ and $1.8 \times 10^{17}$ erg gm$^{-1}$s$^{-1}$, respectively.
At this time the mass fractions of
the positron-decaying nuclei in the nuclear burning region 
now exceed those of most of the stable CNO nuclei ($^{13}$N = $1.7 \times 10^{-2}$,
 $^{14}$O = $1.3 \times 10^{-1}$,$^{15}$O = $2.6 \times 10^{-3}$).  Any further rise
in energy generation will require these nuclei to decay before being able to capture another proton \citep{starrfield_1972_aa, starrfield_1989_aa}. 
Meanwhile, the mass fraction of $^3$He has fallen to $7.6 \times 10^{-6}$ and $^2$H to $4.8 \times 10^{-9}$.  The
mass fraction of $^7$Be has risen to $1.3 \times 10^{-5}$ at the CEI but is only
$8.7 \times 10^{-6}$ at the surface.

At an evolution time of 90.65 s (40 s after peak energy generation) the peak temperature of $3.41 \times 10^8$ K is reached in the mass zones just above
the CEI.  The peak rate of energy generation in the same zone has declined to $3.3 \times 10^{15}$ erg gm$^{-1}$s$^{-1}$, and 
the density (in the same mass zone) to $6.0 \times 10^2$ gm cm$^{-3}$.  The temperatures throughout the nuclear burning region now exceed the Fermi temperature,
lifting electron degeneracy, and the
heating from the nuclear energy release throughout the envelope (the energy generation at the surface now exceeds $3 \times 10^{14}$ erg gm$^{-1}$s$^{-1}$)
has driven the luminosity to $3.1 \times 10^{4}$ L$_\odot$ and T$_{\rm eff}$
to its peak  value of $10^6$K. 

The outer layers are expanding at 20.8 km s$^{-1}$ and the radius of the WD
has increased to 3916 km from $\sim$2700 km.  The expansion velocity at this time is far less than the escape velocity at this radius ($\sim 10^4$ km s$^{-1}$).  
As the outer layers continue their expansion and begin to cool, convection now begins
to retreat from the outer layers and thus the nuclear abundances in the material that will 
eventually be ejected are frozen-in.  The surface abundance of  $^3$He is $4.4 \times 10^{-6}$, that of $^7$Be is
$1.7 \times 10^{-5}$, and $^7$Li is $7.8 \times 10^{-13}$.  The destruction of the initial lithium in this type of evolution is well
understood \citep{cameron_1971_aa} and implies that the $^7$Li and $^7$Be observed in nova ejecta 
\citep{tajitsu_2015_aa, tajitsu_2016_aa, izzo_2015_aa, izzo_2018_aa, molaro_2016_aa, selvelli_2018_aa,wagner_2018_aa} 
must be coming from the decay of $^7$Be produced in the
outburst. 

We continue to evolve the simulation and, after 1.1 hr of expansion, the outer layers are becoming optically thin and their velocities
have reached (because of radiation pressure) $1.4 \times 10^{3}$ km s$^{-1}$.  The surface parameters are:  T$_{\rm eff}$ = $1.3 \times 10^4$K, L = $1.3 \times 10^{5}$ L$_\odot$,
and the outer radius is $5.2 \times 10^{12}$ cm.  At this distance the escape speed has declined to $<$ 200 km s$^{-1}$
so that $4.6 \times 10^{-6}$M$_\odot$ exceeds this speed, is optically thin, and we tabulate it as ejected (see Table \ref{evolCOMDTNR}).
The mass fraction of $^7$Be  is $2.1 \times 10^{-5}$ in the ejected gases.

\section{Nucleosynthesis}
\label{nucleo}

\begin{figure}[htb!]
\includegraphics[width=1.0\textwidth]{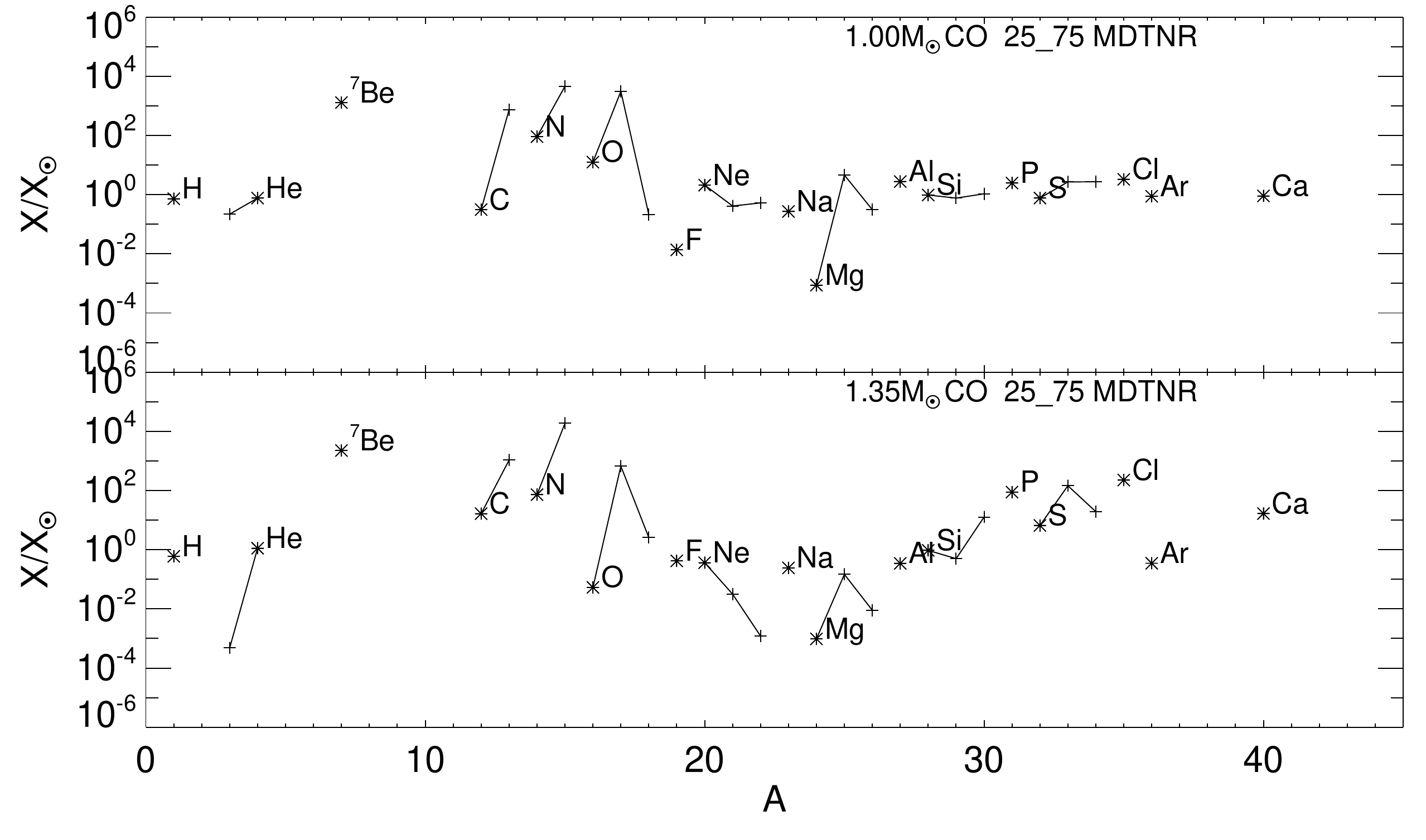}
\caption{Top panel: the abundances of the stable
isotopes from hydrogen to calcium in the ejecta for the
1.0 M$_\odot$ CO WD sequence. The $x$-axis is the atomic mass and the
$y$-axis is the logarithmic ratio of the abundance divided by the
solar abundance \citep{lodders_2003_aa}.  We also include $^7$Be in this plot, even though it is radioactive,  because of its large
overproduction.  Both the initial $^7$Li and $^6$Li are depleted during the evolution.  As in \citet{timmes_1995_aa}, the most abundant isotope of a given element
is designated by an ``$*$'' and all isotopes of a given element
are connected by solid lines.  Any isotope above 1.0 is
overproduced in the ejecta and a number of light, odd isotopes are
significantly enriched in the ejecta as is $^7$Be.
Bottom panel: the same plot as the top panel but for the 1.35 M$_\odot$ simulation with
25\% WD matter and 75\% solar matter.  Because of the higher
peak temperature in this simulation, in addition to the light, odd isotopes of carbon, nitrogen,
and oxygen, phosphorus, and chlorine are also enriched.}
\label{figure2575}
\end{figure}

\begin{figure}[htb!]
\includegraphics[width=1.0\textwidth]{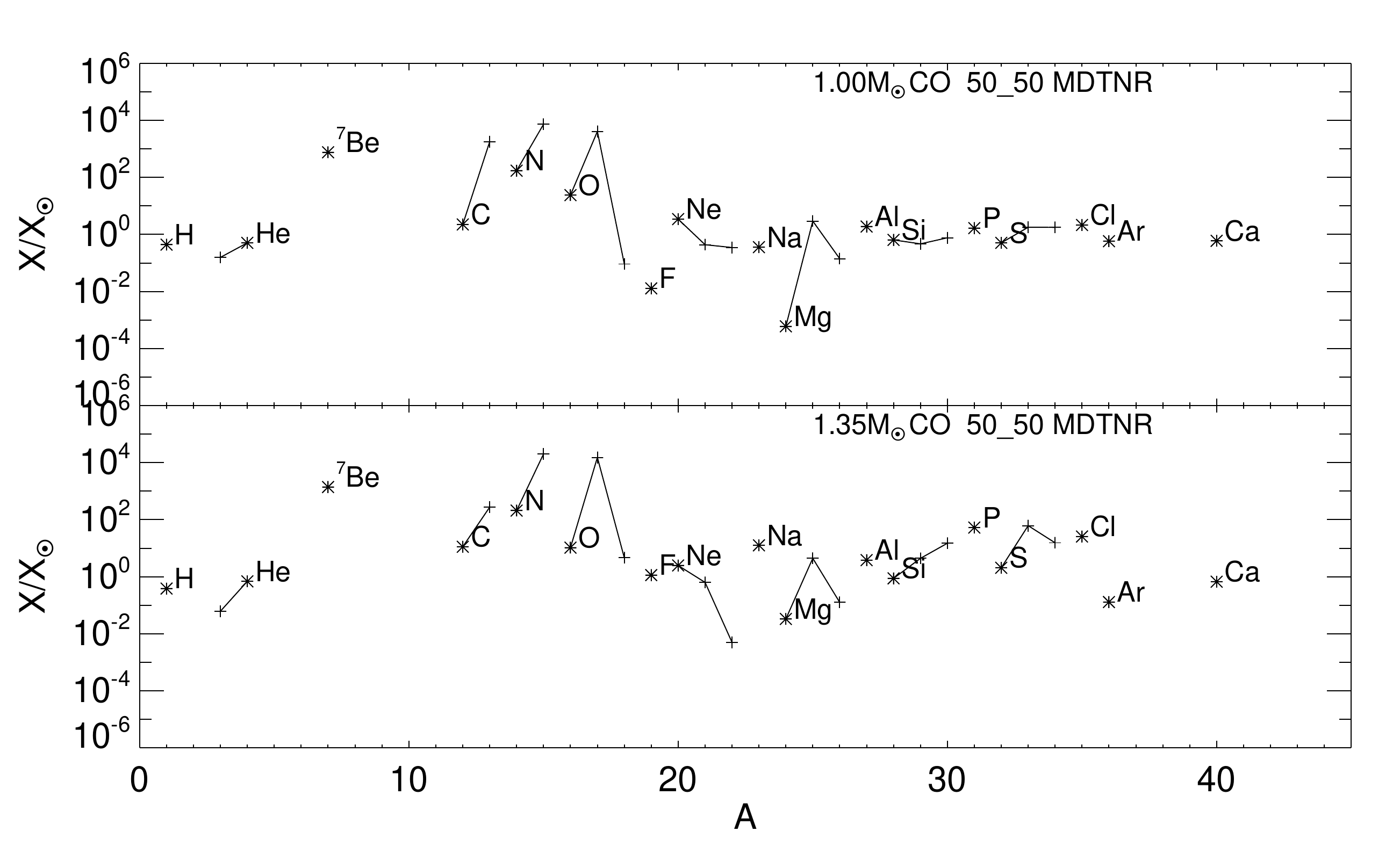}
\caption{Top panel: the same plot as in Figure \ref{figure2575} but for the 1.0 M$_\odot$ 
simulation with 50\% WD matter and 50\% solar matter. The most enriched species are
$^{13}$C, $^{15}$N, $^{17}$O, and $^7$Be. Bottom panel:  the same plot as the bottom panel in Figure \ref{figure2575} but for the simulation with
a mass of 1.35 M$_\odot$ and the 50\% WD and 50\% solar composition.}
\label{figure5050}
\end{figure}

In this section we discuss the nucleosynthesis results from our simulations.  We provide these results both as tables 
of the ejecta abundances in mass fraction and production plots.   Figure \ref{figure2575}  (top panel: 1.0 M$_\odot$ 25\_75 MDTNR; bottom panel: 1.35 M$_\odot$ 25\_75 MDTNR)
and Figure \ref{figure5050} (top panel: 1.0 M$_\odot$ 50\_50 MDTNR; bottom panel: 1.35 M$_\odot$ 50\_50 MDTNR) show the abundances of the stable isotopes
(but also including $^7$Be)  divided by the Lodders (2003) solar
abundances.
In these two figures, the $x$-axis is
the atomic mass number and the $y$-axis is the logarithmic ratio of
the ejecta abundance divided by the solar abundance of the same
isotope.  The most abundant isotope of a given element is marked
by an asterisk and isotopes of the same element are connected by
solid lines and labeled by the given element.  In the next subsection, we present the $^7$Be results and then
follow with a subsection on the other isotopes.

\subsection{The production of $^7$Be in CO classical novae}
\label{lithium}

Because of the recent discoveries of $^7$Be, and its decay product $^7$Li, in CNe ejecta, we 
report in this section that our mixed CO sequences are ejecting amounts of $^7$Be (which decays to $^7$Li after the simulation has ended) 
that are significantly enriched with respect to solar $^7$Li.  

\begin{deluxetable}{@{}lcccccc}
\tablecaption{Comparison of both $^7$Be ejecta and Ejected Mass results with
Jos\'e and Hernanz (1998) and Rukeya et al. (2017)  \label{comparison}}
\tablewidth{0pt}
\tablecolumns{7}
\tablehead{ \colhead{CO WD Mass (M$_\odot$): }&
\colhead{0.8}&
\colhead{0.8}&
\colhead{1.0}&
\colhead{1.15}&
\colhead{1.15}&
\colhead{1.15\tablenotemark{a}}} 

\startdata
Core \%{\tablenotemark{b}}&25&50&50&25&50&50\\
\hline
{\bf $^7$Be ejecta abundance by mass}&&&&&&\\
\hline
\citet{jose_1998_aa}& $4.4 \times 10^{-7}$ & $9.6\times 10^{-7}$&$3.1 \times 10^{-6}$&
$6.0 \times 10^{-6}$&$8.1 \times 10^{-6}$&$3.1 \times10^{-6}$\\
\citet{rukeya_2017_aa}& $5.5 \times 10^{-7}$ & $4.6\times 10^{-7}$&$1.6\times 10^{-6}$&
$4.3 \times 10^{-6}$&$2.9 \times 10^{-6}$&\\
MFB (This Work)&$8.2 \times 10^{-7}$ & $7.0\times 10^{-7}$&$1.4\times 10^{-6}$&
$5.9 \times 10^{-6}$&$4.4 \times 10^{-6}$&\\
MDTNR (This Work)&$3.7 \times 10^{-6}$ & $3.5\times 10^{-6}$&$7.1\times 10^{-6}$&
$1.9 \times 10^{-5}$&$1.2 \times 10^{-5}$&\\
\hline
{\bf Ejected Mass (M$_\odot$)}&&&&&&\\
\hline
\citet{jose_1998_aa}& $7.0 \times 10^{-5}$ & $6.4\times 10^{-5}$&$2.3 \times 10^{-5}$&
$1.5 \times 10^{-5}$&$1.3 \times 10^{-5}$&$6.3 \times10^{-6}$\\
\citet{rukeya_2017_aa}& $2.0 \times 10^{-5}$ & $1.3\times 10^{-5}$&$8.2\times 10^{-6}$&
$4.9 \times 10^{-6}$&$3.6 \times 10^{-6}$&\\
MFB (This Work)&$3.7 \times 10^{-7}$ & $4.1\times 10^{-7}$&$4.4\times 10^{-8}$&
$9.8 \times 10^{-8}$&$1.3 \times 10^{-7}$&\\
MDTNR (This Work)&$2.9 \times 10^{-6}$ & $1.1\times 10^{-4}$&$6.3\times 10^{-5}$&
$1.3 \times 10^{-5}$&$3.4 \times 10^{-5}$&\\
\enddata
\tablenotetext{a}{This sequence is reported on in Table 2 of \citet{jose_1998_aa} and uses the updated opacities of
\citet{iglesias_1993_aa}}
\tablenotetext{b}{The numbers in this row are the percent of core material in the simulation.}
\end{deluxetable}

In Table \ref{comparison} we compare the values in both our MFB and MDTNR studies with those in \citet{hernanz_1996_aa}, \citet{ jose_1998_aa}, and \citet{rukeya_2017_aa}.
\citet{rukeya_2017_aa} also provide a comparison with \citet{jose_1998_aa}.  
Although there are differences between the microphysics in SHIVA \citep{jose_1998_aa} and NOVA (opacities, equations of state, 
nuclear reaction rate library) and in the treatment of convection,  
except for the simulation at 0.6M$_\odot$, there is good agreement in our 2 predictions of $^7$Li  ejecta abundances.  
The agreement is also good comparing our results with \citet{rukeya_2017_aa} who used 
MESA \citep{paxton_2011_aa, paxton_2013_aa, paxton_2015_aa, paxton_2016_aa, paxton_2018_aa} in their study.

The top row lists the WD mass and the next row gives the specific mixture, either
25\% WD matter or 50\% WD matter.  The next set of rows is the comparison of the $^7$Be results from each of the studies listed in the left column. 
The values in the first three rows all  assume MFB.  The results from \citet{ jose_1998_aa} are higher than those of \citet{rukeya_2017_aa} except for that of 25\% WD matter at 
0.8 M$_\odot$.  However, the last column, in which \citet{ jose_1998_aa} redid the same evolutionary sequence, as in the previous column, 
but with the \citet{iglesias_1993_aa} opacities, is nearly identical to that of \citet{rukeya_2017_aa}.  Comparing our MFB simulations to those above, however, we find that 
our $^7$Be predictions exceed those of \citet{rukeya_2017_aa} except for the simulation with 50\% core matter on a 1.0 M$_\odot$ WD.  In contrast, they fall below those
of \citet{ jose_1998_aa} except for the simulations with 25\% core matter at 0.8 M$_\odot$ and their last simulation with the new opacities.  Nevertheless, 
our MDTNR results are always larger than those reported in both the other studies and our MDTNR value for 50\% core matter on a 1.15 M$_\odot$ WD is 
4 times larger than the value reported in \citet{jose_1998_aa} using newer opacities.

We also show in this table the comparison of the amount of ejected mass.  For these cases, the sequences listed for \citet{jose_1998_aa} all eject more mass than
either \citet{rukeya_2017_aa} or our MFB set of calculations.  Once \citet{jose_1998_aa} switch to an updated opacity table, however, their ejected mass drops by a factor of two and 
is more in line with \citet{rukeya_2017_aa}.  Our MFB results are considerably smaller than either of the other two studies.  In tests done to better understand this
difference, we find that the introduction of new electron degenerate conductivities strongly effects the structure of the TNR and reduces the amount
of ejected material.  In addition, \citet{jose_1998_aa} use fewer mass zones ($\sim$ 35) with (probably) larger masses.
However, comparing our MDTNR values for the amount of mass ejected, they are larger than \citet{jose_1998_aa} for the 3 simulations with 50\% core material but
smaller for the 0.8 M$_\odot$ (25\% core matter) and the 1.15$_\odot$ (25\% core matter).  Finally, except for the simulation with 25\% WD matter at 
0.8 M$_\odot$, they are all larger than the equivalent simulations by \citet{rukeya_2017_aa}.

\begin{figure}[htb!]
\includegraphics[width=1.0\textwidth]{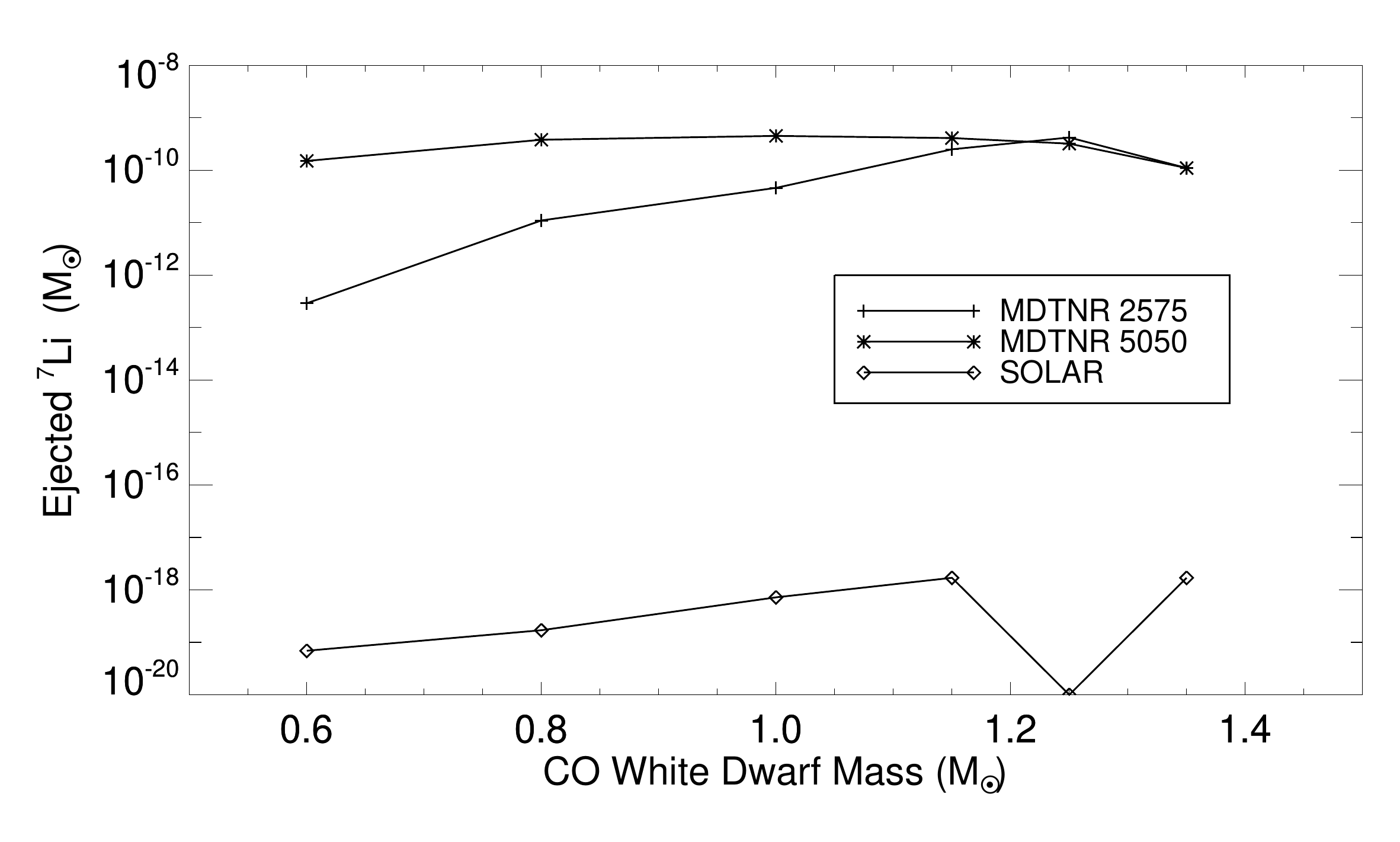}
\caption{The predicted $^7$Li abundance in the ejecta as a function of WD mass in units of solar masses.  The TNRs on CO WDs reach sufficiently 
high temperatures to deplete the initial $^7$Li 
present in the accreted material.  The TNR then produces $^7$Be which is mixed to the surface by strong
convection during the TNR and we actually plot that nucleus.  $^7$Be decays ($\sim$ 53 day half-life) after the end
of the simulations. The simulations where we mix from the beginning (MFB) eject far less $^7$Li 
and are not plotted here. The simulation with solar abundances on a 1.25 M$_\odot$ WD did not eject
any material.}
\label{figureli7mass}
\end{figure}

The amount of $^7$Li (actually produced as $^7$Be) in the ejected material in solar masses is shown in Figure \ref{figureli7mass} as a function of CO WD mass.     All our MDTNR sequences eject material enriched in $^7$Be and 
the amount of enrichment is an increasing function of CO WD mass  \citep{hernanz_1996_aa, jose_1998_aa}.  
The nucleus produced during the TNR is  $^7$Be. However, we do not follow the simulations sufficiently long for $^7$Be to decay to $^7$Li.  
All the initial $^7$Li (or $^6$Li) in the accreting material is destroyed by the TNR.
Both Table \ref{evolCOMFB} and \ref{evolCOMDTNR} give the $^7$Li abundance (assuming that the $^7$Be has decayed) 
as the amount of $^7$Li ejected with respect to the solar value (N($^7$Li/H)$_{\rm ej}$/N($^7$Li/H)$_{\odot}$).  

\begin{deluxetable}{@{}lcccccc}
\tablecaption{Ejecta or Surface Abundances for Solar Accretion and No Mixing with Core Material\tablenotemark{a} \label{cosolarabund}}
\tablewidth{0pt}
\tablecolumns{7}
 \tablehead{ \colhead{WD Mass (M$_\odot$): }&
 \colhead{0.6}&
\colhead{0.8} &
\colhead{1.0} &
\colhead{1.15} &
\colhead{1.25\tablenotemark{b}} &
\colhead{1.35}} 

\startdata
H&$7.0 \times 10^{-1}$&$7.0 \times 10^{-1}$&$7.0 \times 10^{-1}$&$6.7 \times 10^{-1}$&$6.7 \times 10^{-1}$&$6.4 \times 10^{-1}$ \\
$^{3}$He&$7.5\times10^{-12}$  &$3.0 \times 10^{-12}$&$2.0 \times 10^{-7}$& $2.0\times 10^{-8}$&$5.6\times 10^{-13}$ &$8.6 \times 10^{-11}$\\
$^4$He&$2.9 \times 10^{-1}$&$3.0 \times 10^{-1}$&$2.9 \times 10^{-1}$&$3.1 \times 10^{-1}$&$3.2 \times 10^{-1}$&$3.5 \times 10^{-1}$\\
$^{7}$Be&$1.0\times10^{-13}$  &$1.4 \times 10^{-13}$&$1.5 \times 10^{-10}$& $4.5 \times 10^{-11}$&$4.4 \times 10^{-12}$ &$5.7 \times 10^{-11}$\\
$^{7}$Li &0.0 &0.0&$6.3 \times 10^{-11}$& $6.0 \times 10^{-12}$&0.0&$7.4 \times 10^{-15}$\\
$^{12}$C&$1.9\times 10^{-4}$  &$3.4 \times 10^{-4}$&$4.6 \times 10^{-4}$& $8.8 \times 10^{-4}$&$7.2 \times 10^{-4}$ &$1.1 \times 10^{-3}$\\
$^{13}$C &$1.1\times 10^{-4}$ &$3.0 \times 10^{-4}$&$6.0 \times 10^{-4}$& $1.5 \times 10^{-3}$&$1.1 \times 10^{-3}$&$2.1 \times 10^{-3}$\\
$^{14}$N &$7.5\times10^{-3}$ &$7.8 \times 10^{-3}$&$3.3 \times 10^{-3}$& $3.4 \times 10^{-3}$&$3.3 \times 10^{-3}$ &$3.8 \times 10^{-3}$\\
$^{15}$N &$5.9\times10^{-6}$ &$4.2\times10^{-5}$&$4.8\times10^{-3}$& $3.6\times 10^{-3}$&$4.4 \times 10^{-3}$ &$2.3\times10^{-3}$\\
$^{16}$O &$1.9\times10^{-3}$ &$1.1 \times 10^{-3}$&$4.6 \times 10^{-4}$& $3.0 \times 10^{-5}$&$7.2 \times 10^{-6}$ &$1.7 \times 10^{-5}$\\
$^{17}$O&$1.2\times10^{-5}$  &$7.7 \times 10^{-6}$&$1.2 \times 10^{-4}$& $9.4 \times 10^{-6}$&$9.6 \times 10^{-7}$ &$6.1 \times 10^{-7}$\\
$^{18}$O &$5.7\times10^{-9}$ &$3.1\times 10^{-9}$&$1.3 \times 10^{-7}$& $1.8 \times 10^{-8}$&$3.2 \times 10^{-10}$ &$8.1\times 10^{-10}$\\
$^{18}$F &$4.6\times10^{-11}$ &$1.4\times 10^{-10}$&$2.7 \times 10^{-9}$& $6.9 \times 10^{-10}$&$2.3 \times 10^{-11}$ &$5.6\times 10^{-11}$\\
$^{19}$F &$8.1\times10^{-11}$ &$1.3\times 10^{-11}$&$2.9\times 10^{-9}$& $3.1 \times 10^{-10}$&$2.5 \times 10^{-12}$ &$6.3\times 10^{-11}$\\
$^{20}$Ne &$1.2\times10^{-3}$ &$1.2\times 10^{-3}$&$1.2 \times 10^{-3}$& $7.0\times 10^{-4}$&$1.0 \times 10^{-4}$ &$6.0\times 10^{-7}$\\
$^{21}$Ne &$2.3\times10^{-8}$ &$4.0\times 10^{-8}$&$5.9 \times 10^{-7}$& $2.6 \times 10^{-7}$&$4.1\times 10^{-8}$ &$3.0\times 10^{-10}$\\
$^{22}$Ne&$3.6\times10^{-5}$ &$3.8\times 10^{-5}$&$2.4 \times 10^{-5}$& $7.2 \times 10^{-7}$&$4.2\times 10^{-11}$ &$1.6\times 10^{-9}$\\
$^{22}$Na &$2.3\times10^{-6}$ &$9.8\times 10^{-7}$&$1.8 \times 10^{-6}$& $1.9 \times 10^{-6}$&$1.4\times 10^{-7}$ &$3.6 \times 10^{-8}$\\
$^{23}$Na&$6.3\times10^{-6}$ &$3.2\times 10^{-6}$&$5.3 \times 10^{-6}$& $5.7 \times 10^{-6}$&$3.9\times 10^{-7}$ &$1.1\times 10^{-7}$\\
$^{24}$Mg &$5.6\times10^{-8}$ &$3.2\times 10^{-8}$&$3.4 \times 10^{-6}$& $3.7 \times 10^{-7}$&$4.4 \times 10^{-9}$ &$2.8\times 10^{-9}$\\
$^{25}$Mg &$6.0\times10^{-4}$ &$3.3\times 10^{-4}$&$5.6 \times 10^{-5}$& $6.3 \times 10^{-6}$&$3.9 \times 10^{-7}$ &$3.7\times 10^{-8}$\\
$^{26}$Mg&$7.7\times10^{-5}$ &$2.4\times 10^{-5}$&$5.3 \times 10^{-6}$& $4.3 \times 10^{-7}$&$1.6 \times 10^{-8}$ &$2.4\times 10^{-9}$\\
$^{26}$Al &$2.9\times10^{-5}$&$2.0\times 10^{-5}$&$6.5 \times 10^{-6}$& $8.5 \times 10^{-7}$&$1.0 \times 10^{-7}$&$1.6 \times 10^{-9}$ \\
$^{27}$Al &$9.7\times10^{-5}$&$1.4 \times 10^{-4}$&$3.0 \times 10^{-5}$& $4.3\times 10^{-6}$&$4.3 \times 10^{-7}$&$1.9 \times 10^{-8}$ \\
$^{28}$Si &$7.8\times10^{-4}$&$1.1 \times 10^{-3}$&$1.6 \times 10^{-3}$& $7.3 \times 10^{-4}$&$6.6 \times 10^{-5}$ &$5.2 \times 10^{-7}$\\
$^{29}$Si &$4.0\times10^{-5}$&$3.6 \times 10^{-5}$&$2.2 \times 10^{-5}$& $1.3 \times 10^{-5}$&$9.1 \times 10^{-7}$ &$6.0 \times 10^{-8}$\\
$^{30}$Si&$2.7 \times 10^{-5}$&$3.1 \times 10^{-5}$&$8.8\times 10^{-5}$& $6.5 \times 10^{-4}$&$6.5 \times 10^{-5}$&$3.1 \times 10^{-7}$ \\
$^{31}$P &$7.5\times10^{-6}$&$7.2 \times 10^{-6}$&$9.9 \times 10^{-6}$& $8.7 \times 10^{-5}$&$8.2 \times 10^{-6}$ &$1.2 \times 10^{-7}$\\
$^{32}$S &$4.0\times10^{-4}$ &$4.0\times 10^{-4}$&$4.0 \times 10^{-4}$& $1.7 \times 10^{-3}$&$3.9\times 10^{-3}$ &$6.4\times 10^{-4}$\\
$^{33}$S &$3.2\times10^{-6}$&$3.2 \times 10^{-6}$&$2.9 \times 10^{-6}$& $2.3 \times 10^{-6}$&$8.0 \times 10^{-6}$&$2.2 \times 10^{-6}$ \\
$^{34}$S &$1.9\times10^{-5}$ &$1.8\times 10^{-5}$&$1.6 \times 10^{-5}$& $2.5 \times 10^{-6}$&$6.7 \times 10^{-6}$ &$2.4\times 10^{-6}$\\
$^{35}$Cl &$4.1\times10^{-6}$&$4.4\times 10^{-6}$&$6.9 \times 10^{-6}$& $1.7 \times 10^{-5}$&$5.9\times 10^{-5}$ &$4.8 \times 10^{-5}$\\
$^{36}$Ar &$9.1\times10^{-5}$&$9.1 \times 10^{-5}$&$7.8 \times 10^{-5}$& $5.1 \times 10^{-6}$&$5.0 \times 10^{-6}$&$4.2 \times 10^{-6}$ \\
$^{40}$Ca &$7.1\times10^{-5}$&$7.1 \times 10^{-5}$&$7.1 \times 10^{-5}$& $7.2 \times 10^{-5}$&$9.4 \times 10^{-5}$&$3.6 \times 10^{-3}$ \\
\enddata
\tablenotetext{a}{All abundances are Mass Fraction}

\tablenotetext{b}{These are the surface zone abundances since no material was ejected.}

\end{deluxetable}

\begin{deluxetable}{@{}lcccccc}
\tablecaption{Ejecta Abundances for 25-75 MDTNR mixture in
CO White Dwarfs\tablenotemark{a} \label{co2575abund}}
\tablewidth{0pt}
\tablecolumns{7}
 \tablehead{ \colhead{WD Mass (M$_\odot$)}&
 \colhead{0.6}&
\colhead{0.8} &
\colhead{1.0} &
\colhead{1.15} &
\colhead{1.25} &
\colhead{1.35}} 

\startdata
H&$5.3 \times10^{-1}$&$5.2 \times10^{-1}$&$5.1 \times10^{-1}$&$4.8 \times10^{-1}$&$4.6 \times10^{-1}$&$4.3 \times10^{-1}$ \\
$^{3}$He&$3.5\times10^{-5}$  &$1.6 \times 10^{-5}$&$6.4 \times 10^{-6}$& $4.2\times 10^{-7}$&$4.4\times 10^{-8}$ &$1.4 \times 10^{-8}$\\
$^4$He&$2.1 \times10^{-1}$&$2.1 \times10^{-1}$&$2.1 \times10^{-1}$&$2.3 \times10^{-1}$& $2.7 \times10^{-1}$&$3.1 \times10^{-1}$\\
$^{7}$Be&$6.0\times10^{-7}$  &$3.7 \times 10^{-6}$&$1.2 \times 10^{-5}$& $1.9 \times 10^{-5}$&$2.0 \times 10^{-5}$ &$2.1 \times 10^{-5}$\\
$^{7}$Li &$4.8\times10^{-10}$ &$7.1 \times 10^{-13}$&$4.0 \times 10^{-12}$& $4.4 \times 10^{-13}$&$3.1 \times 10^{-13}$&$4.6 \times 10^{-13}$\\
$^{12}$C&$9.0\times 10^{-2}$  &$1.7 \times 10^{-2}$&$9.3 \times 10^{-4}$& $1.0 \times 10^{-2}$&$2.5 \times 10^{-2}$ &$5.0 \times 10^{-2}$\\
$^{13}$C &$3.5\times 10^{-2}$ &$8.2 \times 10^{-2}$&$2.5 \times 10^{-2}$& $1.2 \times 10^{-2}$&$2.1 \times 10^{-2}$&$3.8 \times 10^{-2}$\\
$^{14}$N &$5.2\times10^{-3}$ &$3.8 \times 10^{-2}$&$1.0 \times 10^{-1}$& $8.0 \times 10^{-2}$&$7.9 \times 10^{-2}$ &$8.0 \times 10^{-2}$\\
$^{15}$N &$2.2\times10^{-5}$ &$2.1\times10^{-3}$&$2.1\times10^{-2}$& $9.7\times 10^{-2}$&$1.2\times10^{-1}$&$8.4 \times 10^{-2}$\\
$^{16}$O &$1.3\times10^{-1}$ &$1.3\times10^{-1}$&$1.2\times10^{-1}$& $4.1\times10^{-2}$&$1.4\times10^{-3}$ &$5.4 \times 10^{-4}$\\
$^{17}$O&$2.4\times10^{-4}$  &$2.0 \times 10^{-3}$&$1.3 \times 10^{-2}$& $3.5 \times 10^{-2}$&$1.6 \times 10^{-2}$ &$3.3 \times 10^{-3}$\\
$^{18}$O &$7.3\times10^{-7}$ &$1.8\times 10^{-7}$&$5.3 \times 10^{-6}$& $1.7 \times 10^{-5}$&$3.4 \times 10^{-5}$ &$6.7\times 10^{-5}$\\
$^{18}$F &$1.9\times10^{-9}$ &$1.0\times 10^{-8}$&$6.2 \times 10^{-7}$& $1.6 \times 10^{-6}$&$2.5 \times 10^{-6}$ &$4.3\times 10^{-6}$\\
$^{19}$F &$3.2\times10^{-7}$ &$7.4\times 10^{-9}$&$6.0\times 10^{-9}$& $8.6 \times 10^{-8}$&$3.0 \times 10^{-7}$ &$1.9\times 10^{-7}$\\
$^{20}$Ne &$8.8\times10^{-4}$ &$8.8\times 10^{-4}$&$9.3 \times 10^{-4}$& $1.5\times 10^{-3}$&$1.6 \times 10^{-3}$ &$4.7\times 10^{-4}$\\
$^{21}$Ne &$5.3\times10^{-7}$ &$2.3\times 10^{-7}$&$4.7 \times 10^{-7}$& $6.3 \times 10^{-7}$&$5.8\times 10^{-7}$ &$1.1\times 10^{-7}$\\
$^{22}$Ne&$2.6\times10^{-3}$ &$2.6\times 10^{-3}$&$2.5 \times 10^{-3}$& $1.4 \times 10^{-3}$&$2.6\times 10^{-5}$ &$5.8\times 10^{-6}$\\
$^{22}$Na &$1.8\times10^{-6}$ &$8.6\times 10^{-7}$&$1.4 \times 10^{-6}$& $2.3 \times 10^{-6}$&$5.3\times 10^{-6}$ &$4.4 \times 10^{-6}$\\
$^{23}$Na&$2.9\times10^{-5}$ &$2.6\times 10^{-5}$&$8.8 \times 10^{-6}$& $1.2 \times 10^{-5}$&$2.3\times 10^{-5}$ &$5.9\times 10^{-6}$\\
$^{24}$Mg &$4.1\times10^{-4}$ &$4.7\times 10^{-5}$&$7.2 \times 10^{-7}$& $3.3 \times 10^{-7}$&$3.4 \times 10^{-7}$ &$3.4\times 10^{-7}$\\
$^{25}$Mg &$6.9\times10^{-5}$ &$4.5\times 10^{-4}$&$3.2 \times 10^{-4}$& $7.1 \times 10^{-5}$&$3.4\times 10^{-5}$ &$1.2\times 10^{-5}$\\
$^{26}$Mg&$6.4\times10^{-5}$ &$6.2\times 10^{-5}$&$2.4 \times 10^{-5}$& $3.0 \times 10^{-6}$&$1.7\times 10^{-6}$ &$1.2\times 10^{-6}$\\
$^{26}$Al &$1.6\times10^{-8}$&$3.2 \times 10^{-6}$&$8.4\times 10^{-5}$& $2.2 \times 10^{-5}$&$8.5 \times 10^{-6}$&$2.4 \times 10^{-6}$ \\
$^{27}$Al &$5.0\times10^{-5}$&$5.1 \times 10^{-5}$&$1.7\times 10^{-4}$& $1.2\times 10^{-4}$&$4.2 \times 10^{-5}$&$1.5 \times 10^{-5}$ \\
$^{28}$Si &$5.7\times10^{-4}$&$5.7 \times 10^{-4}$&$6.3 \times 10^{-4}$& $1.6\times 10^{-3}$&$9.1 \times 10^{-4}$ &$4.8 \times 10^{-4}$\\
$^{29}$Si &$3.0\times10^{-5}$&$3.0 \times 10^{-5}$&$2.6 \times 10^{-5}$& $3.0 \times 10^{-5}$&$3.5 \times 10^{-5}$ &$1.7 \times 10^{-5}$\\
$^{30}$Si&$2.0 \times 10^{-5}$&$2.0 \times 10^{-5}$&$2.5 \times 10^{-5}$& $1.3 \times 10^{-4}$&$5.2 \times 10^{-4}$&$2.3 \times 10^{-4}$ \\
$^{31}$P &$5.7\times10^{-6}$&$5.7 \times 10^{-6}$&$5.7 \times 10^{-6}$& $2.4 \times 10^{-5}$&$2.9 \times 10^{-4}$ &$1.5 \times 10^{-4}$\\
$^{32}$S &$3.0\times10^{-4}$ &$3.0\times10^{-4}$ &$3.0\times10^{-4}$ &$3.0\times10^{-4}$ &$2.2\times 10^{-3}$&$2.7 \times 10^{-3}$\\
$^{33}$S &$2.4\times10^{-6}$&$2.4 \times 10^{-6}$&$2.4 \times 10^{-6}$& $2.2 \times 10^{-6}$&$1.7 \times 10^{-5}$&$1.6 \times 10^{-4}$ \\
$^{34}$S &$1.4\times10^{-5}$ &$1.4\times10^{-5}$ &$1.4\times10^{-5}$ &$1.1\times 10^{-5}$&$7.7 \times 10^{-6}$& $1.2 \times 10^{-4}$\\
$^{35}$Cl &$3.0\times10^{-6}$&$3.0 \times 10^{-6}$&$3.2 \times 10^{-6}$& $5.9 \times 10^{-6}$&$1.4\times 10^{-5}$ &$2.7 \times 10^{-4}$\\
$^{36}$Ar &$6.8\times10^{-5}$&$6.8 \times 10^{-5}$&$6.8 \times 10^{-5}$& $4.7 \times 10^{-5}$&$3.0 \times 10^{-6}$&$3.2 \times 10^{-5}$ \\
$^{40}$Ca &$5.4\times10^{-5}$&$5.4 \times 10^{-5}$&$5.4 \times 10^{-5}$& $5.4 \times 10^{-5}$&$5.4 \times 10^{-5}$&$1.0 \times 10^{-3}$ \\
\enddata
\tablenotetext{a}{All abundances are Mass Fraction}

\end{deluxetable}

\begin{deluxetable}{@{}lccccccc}
\tablecaption{Ejecta Abundances for
50-50 MDTNR mixture in CO White Dwarfs\tablenotemark{a} 
\label{co5050abund}}
\tablewidth{0pt}
\tablecolumns{7}
 \tablehead{ \colhead{WD Mass (M$_\odot$): }&
\colhead{0.6} & 
\colhead{0.8} &
\colhead{1.0} &
\colhead{1.15} &
\colhead{1.25} &
\colhead{1.35}} 

\startdata
H&$3.4 \times10^{-1}$&$3.3 \times10^{-1}$&$3.1 \times10^{-1}$&$3.0 \times10^{-1}$&$2.7 \times10^{-1}$&$2.2 \times10^{-1}$ \\
$^{3}$He&$1.1\times10^{-5}$  &$7.6 \times 10^{-6}$&$4.6 \times 10^{-6}$& $1.6 \times 10^{-6}$&$9.3\times 10^{-7}$ &$4.4 \times 10^{-7}$\\
$^4$He&$1.4 \times10^{-1}$&$1.4 \times10^{-1}$&$1.4 \times10^{-1}$&$1.6 \times10^{-1}$&$1.9\times10^{-1}$&$2.4 \times10^{-1}$\\
$^{7}$Be&$9.3\times10^{-7}$  &$3.5 \times 10^{-6}$&$7.1 \times 10^{-6}$& $1.2 \times 10^{-5}$&$1.4\times 10^{-5}$ &$1.6 \times 10^{-5}$\\
$^{7}$Li  &$4.3\times10^{-13}$&$2.1 \times 10^{-13}$&$1.2\times 10^{-13}$& $1.2\times 10^{-13}$&$3.0\times 10^{-13}$&$7.2 \times 10^{-14}$\\
$^{12}$C &$1.0 \times10^{-1}$&$2.3\times10^{-2}$&$6.6 \times 10^{-3}$& $1.7 \times 10^{-2}$&$2.0 \times 10^{-2}$ &$3.6\times 10^{-2}$\\
$^{13}$C &$1.3 \times10^{-1}$&$1.5 \times10^{-1}$&$6.5\times10^{-2}$ &$1.2 \times 10^{-2}$&$2.1 \times 10^{-2}$&$1.0 \times 10^{-2}$\\
$^{14}$N&$3.0\times10^{-2}$ &$9.6 \times 10^{-2}$&$1.9 \times10^{-1}$&$2.0 \times10^{-1}$&$1.9 \times10^{-1}$&$1.7 \times10^{-1}$\\
$^{15}$N &$2.2\times10^{-4}$ &$5.3\times10^{-3}$&$3.2\times10^{-2}$& $8.8\times 10^{-2}$&$1.7 \times10^{-1}$&$2.0 \times 10^{-1}$ \\
$^{16}$O &$2.5 \times10^{-1}$ &$2.4 \times10^{-1}$&$2.3 \times10^{-1}$&$1.8 \times10^{-1}$ &$4.9\times10^{-2}$&$1.9 \times10^{-2}$\\
$^{17}$O &$1.4\times10^{-3}$ &$4.9 \times 10^{-3}$&$1.6 \times 10^{-2}$& $4.3 \times 10^{-2}$&$8.6 \times 10^{-2}$ &$3.2 \times 10^{-2}$\\
$^{18}$O &$3.9\times10^{-7}$ &$8.2\times 10^{-7}$&$2.0 \times 10^{-6}$& $8.3 \times 10^{-6}$&$3.0\times 10^{-5}$ &$6.0\times 10^{-6}$\\
$^{18}$F &$2.1\times10^{-8}$ &$4.3\times 10^{-8}$&$5.0 \times 10^{-7}$& $4.0 \times 10^{-6}$&$9.4 \times 10^{-6}$ &$1.9\times 10^{-6}$\\
$^{19}$F &$2.6\times10^{-8}$ &$8.6\times 10^{-10}$&$5.2 \times 10^{-9}$& $4.2 \times 10^{-8}$&$7.3 \times 10^{-7}$ &$1.0\times 10^{-6}$\\
$^{20}$Ne &$5.8\times10^{-4}$ &$6.0\times 10^{-4}$&$6.7 \times 10^{-4}$& $1.1\times 10^{-3}$&$2.6 \times 10^{-3}$ &$1.3\times 10^{-3}$\\
$^{21}$Ne &$3.2\times10^{-8}$ &$1.1\times 10^{-7}$&$2.2 \times 10^{-7}$& $3.8 \times 10^{-7}$&$1.1\times 10^{-6}$ &$4.0\times 10^{-7}$\\
$^{22}$Ne&$5.0\times10^{-3}$ &$5.0\times 10^{-3}$&$4.9 \times 10^{-3}$& $4.0 \times 10^{-3}$&$4.2\times 10^{-4}$ &$1.9\times 10^{-4}$\\
$^{22}$Na &$1.2\times10^{-6}$ &$2.9\times 10^{-7}$&$4.1 \times 10^{-7}$& $1.1 \times 10^{-6}$&$5.1\times 10^{-6}$ &$1.2\times 10^{-6}$\\
$^{23}$Na&$2.2\times10^{-5}$ &$2.1\times 10^{-5}$&$1.3 \times 10^{-5}$& $1.3 \times 10^{-5}$&$4.1\times 10^{-5}$ &$2.2\times 10^{-5}$\\
$^{24}$Mg &$1.3\times10^{-4}$ &$1.1\times 10^{-6}$&$6.4\times 10^{-7}$& $7.1 \times 10^{-7}$&$1.4 \times 10^{-6}$ &$2.4\times 10^{-6}$\\
$^{25}$Mg &$2.0\times10^{-4}$ &$3.2\times 10^{-4}$&$2.4 \times 10^{-4}$& $1.8 \times 10^{-4}$&$2.3 \times 10^{-4}$ &$8.4\times 10^{-5}$\\
$^{26}$Mg&$4.2\times10^{-5}$ &$3.7\times 10^{-5}$&$1.2 \times 10^{-5}$& $9.1 \times 10^{-6}$&$2.0 \times 10^{-5}$ &$6.8\times 10^{-6}$\\
$^{26}$Al &$1.8\times10^{-7}$&$1.2 \times 10^{-5}$&$6.6 \times 10^{-5}$& $7.7 \times 10^{-5}$&$2.9 \times 10^{-5}$&$2.8 \times 10^{-5}$ \\
$^{27}$Al &$3.3\times10^{-5}$&$3.9 \times 10^{-5}$&$1.2 \times 10^{-4}$& $3.0 \times 10^{-4}$&$1.4 \times 10^{-4}$&$1.3 \times 10^{-4}$ \\
$^{28}$Si &$3.8\times10^{-4}$&$3.8 \times 10^{-4}$&$4.2 \times 10^{-4}$& $7.7 \times 10^{-4}$&$2.1 \times 10^{-3}$ &$9.9 \times 10^{-4}$\\
$^{29}$Si &$2.0\times10^{-5}$&$2.0 \times 10^{-5}$&$1.6 \times 10^{-5}$& $1.3 \times 10^{-5}$&$9.2 \times 10^{-5}$ &$3.3 \times 10^{-5}$\\
$^{30}$Si &$1.4\times10^{-5}$&$1.4 \times 10^{-5}$&$1.8 \times 10^{-5}$& $4.0 \times 10^{-5}$&$7.9 \times 10^{-4}$&$4.6 \times 10^{-4}$ \\
$^{31}$P &$3.8\times10^{-6}$&$3.8 \times 10^{-6}$&$3.8 \times 10^{-6}$& $5.5 \times 10^{-6}$&$3.0 \times 10^{-4}$ &$2.9 \times 10^{-4}$\\
$^{32}$S &$2.0\times10^{-4}$&$2.0 \times 10^{-4}$&$2.0 \times 10^{-4}$& $2.0 \times 10^{-4}$&$4.9 \times 10^{-4}$&$3.4 \times 10^{-3}$ \\
$^{33}$S &$1.6\times10^{-6}$&$1.6 \times 10^{-6}$&$1.6 \times 10^{-6}$& $1.6 \times 10^{-6}$&$4.0 \times 10^{-6}$&$2.4 \times 10^{-4}$ \\
$^{34}$S &$9.3\times10^{-6}$&$9.3 \times 10^{-6}$&$9.2 \times 10^{-6}$& $8.7 \times 10^{-6}$&$4.4 \times 10^{-6}$&$1.4 \times 10^{-4}$ \\
$^{35}$Cl &$2.0\times10^{-6}$&$2.0 \times 10^{-6}$&$2.1 \times 10^{-6}$& $2.7 \times 10^{-6}$&$6.8 \times 10^{-6}$ &$2.5 \times 10^{-4}$\\
$^{36}$Ar &$4.6\times10^{-5}$&$4.6 \times 10^{-5}$&$4.5 \times 10^{-5}$& $4.0 \times 10^{-5}$&$1.1 \times 10^{-5}$&$4.8 \times 10^{-5}$ \\
$^{40}$Ca &$3.6\times10^{-5}$&$3.6 \times 10^{-5}$&$3.6 \times 10^{-5}$& $3.6 \times 10^{-5}$&$3.6 \times 10^{-5}$&$5.9 \times 10^{-5}$ \\
\enddata
\tablenotetext{a}{All abundances are Mass Fraction}

\end{deluxetable}

\subsection{Enrichment of the other Nuclei in CO novae ejecta}

Figures \ref{figure2575}  and \ref{figure5050} show for both WD masses and compositions that $^7$Be, 
$^{15}$N, $^{17}$O,  $^{31}$P, $^{35}$Cl, and $^{40}$Ca are
significantly overproduced in CN ejecta. The results for the 1.0 M$_\odot$ sequences are given in the top panels of 
Figures  \ref{figure2575}  and \ref{figure5050} and they show that both $^7$Be and $^{13}$C are about 300 times solar and 
$^{15}$N and $^{17}$O are nearly $10^4$ times solar.  In contrast, both $^{18}$O and $^{18}$F are depleted. 
None of the other isotopes are significantly enriched in the 1.0M$_\odot$ sequences.   The 1.35 M$_\odot$ results are shown in the bottom
panels of Figure \ref{figure2575} and Figure \ref{figure5050}. 
Table \ref{evolCOMDTNR} shows that peak temperatures in the 50\% WD - 50\% solar sequences 
are much higher than in the 25\% WD - 75\% solar sequences.  Thus, $^{13}$C, $^{15}$N, $^{17}$O, $^{29}$S, $^{31}$P,
and $^{35}$Cl are a great deal more enriched in the 50\% WD - 50\% solar sequence.  In addition, $^7$Be is enriched by about a factor of 300 and $^{22}$Ne is depleted as is $^{23}$Na.  

Tables \ref{cosolarabund}, \ref{co2575abund}, and  \ref{co5050abund} provide the detailed isotopic abundances in the 
ejected matter and allow us to compare the results for different WD masses. We do not include similar tables for the MFB simulations since they
hardly ejected any material. 
Table  \ref{cosolarabund} (no mixing of accreted with core material, hence, a solar mixture only) allows us to make predictions for those CNe or RNe that do not mix with WD matter.
 It shows that the ejected $^{12}$C abundance increases with WD mass, while $^{14}$N is relatively constant and the $^{16}$O abundance declines with increasing WD mass.  
 The odd isotopes, such
 as  $^{13}$C, increase with CO WD mass.  For WD masses that exceed 1.0 M$_\odot$ the $^{13}$C abundance 
 always exceeds that of $^{12}$C.  The $^{15}$N abundance increases with CO WD mass and for some ranges in 
 WD mass (1.0 M$_\odot$ to 1.25 M$_\odot$) its abundance exceeds that of $^{14}$N.  In contrast, $^{18}$O, $^{26}$Al, and $^{27}$Al  
 decrease with increasing CO WD mass.  The abundance  of $^4$He increases with WD mass, implying that more hydrogen is burned to 
 helium to produce the energy radiated by the outburst, since the amount of accreted mass declines with increasing WD mass.
 
Table  \ref{co2575abund} gives the ejecta abundances for the mixture with 25\% WD matter and 75\% solar matter.  The abundance of $^7$Be clearly increases with increasing WD mass and 
the initial $^7$Li is destroyed by the TNR.  Both $^{12}$C and $^{13}$C are produced in the higher mass CO WDs but there is more $^{12}$C than $^{13}$C produced for most WD masses. 
The abundance of $^{14}$N is roughly constant for the more massive CO WDs while $^{15}$N reaches a peak abundance 
of 0.12 for a 1.25 M$_\odot$ WD and is nearly that value for the other massive WDs.  
Moreover, its abundance exceeds that of $^{14}$N for WD masses from 1.15 M$_\odot$ to 1.35 M$_\odot$.   In contrast to the solar abundance results (Table \ref{cosolarabund}), the abundance of
$^{17}$O, $^{18}$O, and $^{31}$P increase with WD mass.   $^{26}$Al and $^{27}$Al reach a 
maximum abundance at 1.0M$_\odot$ and then decline with increasing WD mass as the peak temperature in the nuclear burning region increases 
during the TNR.  The ratio of their abundances is $\sim$0.2.

Table \ref{co5050abund} provides a listing of the ejecta abundances for the mixture with 50\% WD matter and 50\% solar matter.  Again, the ejected hydrogen abundance declines with 
increasing WD mass because the total envelope mass decreases with increasing WD mass so that it takes more hydrogen burning to provide the energy observed in the outburst.  The $^7$Be
abundance reaches a maximum at 1.25 M$_\odot$ but the $^7$Li abundance decreases with increasing WD mass. $^7$Li is essentially destroyed in the outburst 
so that, again, all the $^7$Li observed in CNe ejecta must be coming from the decay of $^7$Be produced in the outburst.   

The ejecta abundance of $^{12}$C increases with WD mass while that of $^{13}$C is maximum at 0.8 M$_\odot$ and then declines.  The abundance of $^{14}$N increases with WD mass while $^{15}$N increases and reaches a maximum at 1.25 M$_\odot$.
In fact, the odd isotopes are so abundant that molecular studies of CN ejecta should discover large amounts of $^{12}$C$^{15}$N,  $^{13}$C$^{14}$N, and in some cases
$^{13}$C$^{15}$N.  The detection of these molecular species would provide strong observational support for the results of these simulations, and possibly could
be used to determine the composition of the underlying WD. 

The abundance of $^{16}$O declines with increasing WD mass
while that of $^{17}$O increases but reaches a maximum value at 1.25 M$_\odot$. In contrast, that of $^{18}$O increases as the WD mass increases.  The abundance of  $^{26}$Al reaches
a maximum at 1.0 M$_\odot$ while that of $^{27}$Al increases up to 1.35 M$_\odot$.  The ratio of their abundances varies from about 0.3 down to about 0.1, values which are smaller than
found in the 25\% WD - 75\% solar MDTNR studies.  The abundances of $^{22}$Na, $^{31}$P and $^{35}$Cl also increase with CO WD mass.  

\medskip
\section{Could CNe and CVs be progenitors of SN Ia?}
\label{progenitors}

Although of great importance to both galactic chemical evolution and, in addition, as probes of  
the evolution of the universe, the progenitors of SN Ia explosions are as yet unknown. 
Originally, the single-degenerate scenario, with the WD accreting from the secondary and growing in mass toward the Chandrasekhar Limit,
was preferred but this scenario is now disfavored by many  \citep[][and references therein]{gilfanov_2010_aa}.  The other scenario, the double-degenerate
scenario, which involves either a merger or collision between two CO WDs, is now thought to be the major channel for SN Ia explosions. 
 The cause of this switch in the preferred explosion paradigm is a number of perceived problems with the single-degenerate scenario that need to be understood.  
 In this section  we discuss
four of those problems and show that they are, in fact, not problems at all.

The first major problem, which is directly relevant to the simulations presented in the earlier sections,
is the common assumption, based on the analyses of the ejecta abundances and ejecta masses of
CNe outbursts that the WD is decreasing in mass as a consequence of the TNR and resulting explosion.  As we have now shown in earlier sections,
however, that assumption is incorrect and, in fact,  the WD in a {\it CO nova outburst} is gaining in mass. 

We have also
shown in previous studies, in addition, that the WD is growing in mass when there is no mixing of the accreting material with WD core matter 
as may be occurring in CVs in general \citep[][and references therein]{starrfield_2014_aa, starrfield_2017_aa}. These latter results
are based on studies with both NOVA and MESA and
imply that the consequence of mass transfer in CVs is the increasing mass of the WD.  
Moreover, the calculations of \citet{hillman_2015_aa, hillman_2016_aa} show that high mass accretion rates also result in the
WDs growing in mass.   
One concern, nevertheless, is that the large number of CVs in the galaxy may
result in too many SN Ia explosions.  We note, however, that the mass of the secondary also determines the 
ultimate consequences of the evolution.
It is possible that in many CV systems the secondary has too little mass and, therefore,
the evolution will end before the WD has reached the Chandrasekhar Limit.

The second perceived problem is due to the {\it interpretation} of the calculations of \citet{nomoto_1982_aa} and \citet{ fujimoto_1982_aa, fujimoto_1982_ab}. 
A reproduction of their results can be found as Figure 5 in \citet{kahabka_1997_aa}. 
The figure shows that the space describing the consequences of mass accretion rate as a function of the mass of the
accreting WD can be divided into three regions.  
For the lowest mass accretion rates, at all WD masses, it is predicted that accretion results in hydrogen flashes that
resemble those of CNe and, as already discussed, the WD is thought to be losing mass.  
However,  the purpose of this paper has been to provide a broad range of simulations at low \.M
that show a WD accreting at low rates is gaining in mass.  \citet{hillman_2015_aa, hillman_2016_aa} have investigated
the consequences of accretion at higher rates and also find that the WD is growing in mass.  Thus, mass accreting systems 
with a broad range in WD mass and  \.M must be included in the classes of SN Ia progenitors.

A third problem relates to the upper region on the \citet{nomoto_1982_aa} and the \citet{fujimoto_1982_aa, fujimoto_1982_ab} plot, which shows the results for the highest
accretion rates and predicts for all WD masses
that the radius of the WD will grow rapidly to red giant dimensions, accretion will be halted, and any further evolution will
await the collapse of the extended layers. These systems, therefore, cannot be SN Ia progenitors.   
However, we have done extensive studies of solar accretion onto WDs using both
NOVA and MESA and our version of their diagram can be found in \citet[][]{starrfield_2014_aa, starrfield_2017_aa}.
Our fully hydrodynamic studies show,
for the highest mass accretion rates on the most massive WDs, steady hydrogen burning  (see below) is occurring followed by recurring helium
flashes.  The helium flashes do not result in ejection and the WDs are again growing in mass. \citet{hillman_2015_aa, hillman_2016_aa} also
report that helium flashes do not eject material.

The fourth problem is based on the existence of the intermediate regime identified by \citet{nomoto_1982_aa} and \citet{fujimoto_1982_aa, fujimoto_1982_ab}, 
where the material is predicted to burn steadily at the rate it is accreted.  The central \.M of this region is nominally 
$\sim 3 \times 10^{-7}$M$_\odot$ yr$^{-1}$ and it does have a slight variation with WD mass.   
Those systems that are accreting at the steady nuclear burning rate are supposedly evolving horizontally in this plot towards higher WD mass 
and, by some {\it unknown} mechanism, the mass transfer in the binary system is stuck in this mass accretion range.  
\citet{vandenheuvel_1992_aa} predicted that it was only the systems in this region that
were SN Ia progenitors via the SD scenario, They identified the Super Soft X-ray sources (SSS) as those systems, based on their luminosities and effective temperatures. 
The SSS  are $\it luminous$, massive, WDs  discovered by ROSAT \citep{Trumper_1991_aa}.  
They are binaries, with luminosities L$_* \sim 10^{37-38}$erg s$^{-1}$
and effective temperatures ranging from $3 - 7 \times 10^5$K \citep{cowley_1998_aa}.  See also \citet{branch_1995_aa} and \citet{kahabka_1997_aa}.  

However, in more recent studies of accretion without mixing, an expanded study of the stability of thin shells can be found in \citet[][and references therein]{yoon_2004_aa}, who investigated
the accretion of helium-rich and hydrogen-rich material onto WDs.  Using their results, we find that sequences in the steady nuclear burning regime begin in their stable region, 
but with continued accretion, evolve into instability.   In addition, their study shows that the evolutionary sequences
at these \.M exhibit the \citet{schwarzschild_1965_aa} thin shell instability, 
which implies that steady burning does not occur.  We identify these systems, therefore, with those CVs (dwarf, recurrent, symbiotic
novae) that show no core material either on the surface of the WD or in their ejecta. 

Given that the SSS were the 
only systems that were predicted to be single-degenerate Ia progenitors, it was expected that they would be detected by consequences of the long
periods of luminous X-ray and UV emission on the surrounding ISM.  In addition, this extreme emission should still be evident in the ISM surrounding recent SN
Ia explosions.   As an example, we quote from \citet{graur_2019_aa} "For the WD to efficiently grow in mass, the accreted hydrogen must undergo stable nuclear-burning on its surface. This means the progenitor system will be a luminous source of soft X-ray emission \citep[a supersoft X-ray source, SSS,][]{vandenheuvel_1992_aa} for at least some period of time before the explosion."  Similar statements can also be found in \citet{gilfanov_2010_aa} and \citet{kuuttila_2019_aa}. 
Such emission has not been found and the absence of evidence has been used to eliminate the
single degenerate scenario even in the most recent studies.  However, observations of CNe and CVs, which we now identify as
possible SN Ia progenitors, show that they do not spend a large amount of time at high luminosities and effective temperatures.  

Moreover, some RNe are repeating sufficiently often that their WDs
must have grown in mass so that they are now close to the Chandrasekhar Limit.  One such system is the ``rapidly recurring'' RN in M31 (M31N 2008-12a) which is outbursting about once per year and
has opened up a large cavity in the ISM surrounding the system \citep{darnley_2016_aa, darnley_2017_aa, darnley_2017_ab, darnley_2019_aa, henze_2015_aa, henze_2018_aa}.  
It is neither X-ray nor UV luminous between outbursts.

\section{Discussion}

 Fortunately for this study, the recent multi-d studies of convection in the accreted layers of WDs \citep[][and references therein]{casanova_2010_aa,
 casanova_2010_ab, casanova_2011_aa, casanova_2011_ab, casanova_2016_aa, casanova_2018_aa, jose_2014_aa} implied that we could reasonably approximate
 their results by accreting a hydrogen-rich (solar abundances) layer and then switch to a mixed composition once the TNR was underway and convection had begun.
A similar technique has already been used by \citet{jose_2007_aa} who explored a variety of time scales for mixing the WD material into the accreted layers, once convection was
underway, and found that using short time scales was warranted.   We chose an ``instantaneous'' time for mixing
both because it is reproducible and because it is not in disagreement with their results.  Moreover, our initial MDTNR studies suggested that CO WDs were growing in mass and we extended our studies to 1.35M$_\odot$ CO WDs. 

Therefore, we have used NOVA to study the consequences of TNRs on WDs of various masses using three different compositions.  In all cases we find that more mass is
accreted than ejected and, therefore, the WD is growing in mass. We have used two different techniques to treat the  accreting material.  In the first we assumed that the
solar material is mixed from the beginning of accretion (MFB).  This is the technique used both by us and others in the past because there was no agreement on
when and how WD material was mixed up into the accreting matter.  Neither the consequences of our solar mixture accretion simulations nor those where we mix from the 
beginning of accretion (MFB) (Tables \ref{evolCOMFB} and  \ref{evolCOMDTNR}) agree with the 
observations of CNe outbursts.  

Switching to a mixed composition once the TNR is ongoing and a major fraction of the accreted material is 
convective, however, provides a range of model outcomes that are more compatible with observed CNe physical parameters 
reported in the literature.  The simulations with 25\% WD matter and 75\% solar matter (MDTNR) appear to fit the observations
somewhat better than those with 50\% WD matter and 50\% solar matter (MDTNR).  Nevertheless, NOVA  is able to only follow one
 outburst and reaching to close to the Chandrasekhar Limit requires many such cycles of accretion-TNR-ejection - accretion.
 While this has yet to be done with either CO or ONe enriched material (this may have been done in the study of
 \citet{rukeya_2017_aa} but they only reported their ejected mass not the accreted mass), multi-cycle evolution and the growth in mass of the
 WD has been done with solar accretion studies \citep{starrfield_2014_aa, hillman_2015_aa, hillman_2016_aa, starrfield_2017_aa}.  We note
 that the multi-cycle studies reported in \citet{starrfield_2014_aa} and \citet{starrfield_2017_aa} were done with 
 MESA \citep[][and references therein]{paxton_2011_aa, paxton_2013_aa, paxton_2018_aa} while those described by
 \citet{hillman_2015_aa} and \citet{hillman_2016_aa} were done with the code of \citet[][and references therein]{kovetz_2009_aa}. 
 Given these studies with multiple codes, therefore, we feel that our single outburst result implies that the consequences of the
 CN outburst is the growth in mass of the WD under all situations.
 
 Of great importance, some of the ejected isotope abundances in the simulations also fit the isotopic ratios measured for some
 pre-solar grains suggesting that these grains come from CNe ejecta \citep{bose_2019_aa}. 
 \citet{bose_2019_aa} compared the compositions of 30 pre-solar SiC grains with the ejected isotopic abundances in 
Tables \ref{co2575abund} and \ref{co5050abund}. 
The simulations with 25\% WD matter and 75\% solar matter with CO WD masses from 0.8 M$_\odot$ to 1.35 M$_\odot$ 
provide the best fits to the measured isotopic data in four SiC grains.  In addition, one grain matches the 50\% WD and 50\% solar 1.35 M$_\odot$ MDTNR simulation.  
To the best of our knowledge, this is the first study that successfully applies CO nova simulations to both observations of nova dust and pre-solar grains of nova origin.
 Previous studies that attempted to understand SiC nova grain candidates used ONe nova simulations from \citet{jose_1998_aa}, which mandated mixing $>$95\% of solar matter with $<$5\% of CN ejected matter to account for the grains' compositions \citep{amari_2001_aa}. Such mixing is not required for other grain types (e.g., SiC X grains from supernovae). The other assumption, that the binary companion to the WD had to be a main sequence star, made the assignment of nova candidate grains as \textit{bona fide} nova grains even more uncertain.  However, using the simulations described here, we require less than 25\% of solar system material be mixed with the CO nova ejecta to account for the grain compositions \citep{bose_2019_aa}.

 Our simulations show that for CO WD mass $\ge$ 1.15 M$_\odot$ the mass fraction of $^7$Li ($^7$Be) ejected is either $2 \times 10^{-5}$ 
 (25\% WD matter and 75\% solar matter: Table \ref{co2575abund}) or $10^{-5}$ (50\% WD matter plus 50\% solar matter: Table \ref{co5050abund}). 
 The amount of ejected mass for the
 same WD range is $\sim 10^{-5}$M$_\odot$ for the 25\% WD matter and 75\% solar matter simulations and $\sim 2 \times 10^{-5}$ for the
 50\% WD matter plus 50\% solar matter simulations as given in Table \ref{evolCOMDTNR}.  Interestingly, their product implies an ejected $^7$Li mass of $\sim 2 \times 10^{-10}$ M$_\odot$ for
 either composition.  If we take a value for the CN rate of 50 yr$^{-1}$ \citep{shafter_2017_aa}, a lifetime for the galaxy of $10^{10}$yr, and our
 production values we arrive at a predicted abundance of $\sim$100M$_\odot$ for the $^7$Li produced by CNe in the galaxy.   

Our results confirm that CO novae are overproducing $^7$Be, which decays to $^7$Li.  
The amount of $^7$Be we predict from our simulations, in combination with the observations,
allow us to assert that CNe are responsible for a significant fraction of the $^7$Li in the galaxy.  
Moreover, the observations of $^7$Be and $^7$Li found in the early high dispersion optical spectra of the ejected material from CN outbursts  (both CO and ONe)
\citep{tajitsu_2015_aa, tajitsu_2016_aa, izzo_2015_aa, izzo_2018_aa, molaro_2016_aa, selvelli_2018_aa,wagner_2018_aa} report much higher
values than we predict.  In fact, at least 10 times higher than previously predicted \citep{starrfield_1978_aa,
hernanz_1996_aa, jose_1998_aa}.

We also address the question: {\it what is the total amount of $^7$Li in the galaxy?}  The number usually quoted is $\sim$150M$_\odot$
 \citep{hernanz_1996_aa, molaro_2016_aa}.  However, we arrive at a different value.  \citet{lodders_2009_ab} give a value of $2.0 \times 10^{-9}$ for the solar system abundance
 of $^7$Li/H by number.  We convert to mass fraction by multiplying by 7 and obtain $1.4 \times 10^{-8}$ for X($^7$Li)/X(H).   We assume that the total mass of the
 galaxy is $\sim 10^{11}$ M$_\odot$ and the mass fraction of hydrogen is 0.71 \citep{lodders_2009_aa,lodders_2009_ab}. Therefore,
 the total mass of $^7$Li in the galaxy should be $0.71 \times 10^{11}$$ \times$$ 1.4 \times 10^{-8}$ or $\sim$1000M$_\odot$. 
The most recent discussion of the importance of CNe for $^7$Li in the galaxy is that of
\citet[][and references therein] {cescutti_2019_aa} who address the discoveries of $^7$Li and $^7$Be in CN
 ejecta.  Finally, the primordial $^7$Li abundance in the galaxy is $\sim$80M$_\odot$ requiring a galactic source of $^7$Li \citep{fields_2011_aa}. 
 $^6$Li is produced by spallation and not by nuclear reactions in stars, however, so that there should not be a correlation in the
 abundances of these two isotopes in stellar sources.   

Our technique, of first accreting a solar mixture and then switching to a mixed composition, can be compared to calculations
where a mixed composition was used from the beginning of the simulation \citep{hernanz_1996_aa, jose_1998_aa, rukeya_2017_aa}.
They accreted onto both CO and ONe WDs in order to determine the $^7$Li production
from CNe but did not study CO WDs as massive as in this work.  
They assumed two mixed compositions from the beginning
(either 25\% WD material or 50\% WD material), with a solar \citep{lodders_2009_aa} $^3$He abundance of $8.46 \times 10^{-5}$.   
All our simulations used a $^3$He mass fraction \citep{lodders_2003_aa} of $3.41 \times 10^{-5}$.  Since the production of $^7$Be occurs through the
$^3$He($\alpha$,$\gamma$)$^7$Be reaction, 
a higher abundance of $^3$He is expected to result in a higher $^7$Be abundance \citep{hernanz_1996_aa, jose_1998_aa}.  However,  the larger
accreted mass in our simulations resulted both in a higher peak temperature
and also stronger convection (transporting the $^7$Be more rapidly to the surface layers).  These effects combined 
resulted in a larger amount of $^7$Be than reported in either  \citet{hernanz_1996_aa}, \citet{jose_1998_aa}, or \citet{rukeya_2017_aa}.

\citet{rukeya_2017_aa} also compared their simulations to the total observed amount of $^7$Li in the galaxy \citep[$\sim$150 M$_\odot$ given in][]{hernanz_1996_aa}.
They argued that CO novae are producing about 10\% of galactic lithium.   Our simulations, however, produce at least 1.5 times 
more $^7$Li than their simulations  \citep[or those of][]{jose_1998_aa}.  In addition, we have also followed the $^7$Li production on 
more massive WDs achieving about a factor of 2 enrichment over their results.  Therefore, it seems likely that CO novae can produce a significant amount of
stellar $^7$Li.  In Tables \ref{evolCOMFB} and \ref{evolCOMDTNR} we give the ejected $^7$Li abundance in the same units as in \citet{hernanz_1996_aa} so that a direct
comparison can be made. 

We find that the amount of {\it accreted} material is an inverse function of the initial abundance of $^{12}$C.  
Accreting solar material (rather than mixed) allows for more matter to be accreted.  Reducing the metallicity to values seen in the LMC, SMC, 
or even lower also reduces the initial $^{12}$C allowing more material to be accreted before the
TNR is initiated \citep{starrfield_1999_ac, jose_2007_aa} and the accreted material mixes with WD matter.  
Finally, either no mixing with the WD (RNe) or mixing too early with the CO WD (MFB) results in an outburst that ejects
less material than is accreted and the WD is also growing in mass.  

Finally, there is little to no observational evidence for mixing of accreted matter with WD matter in RN  explosions.  While all CNe are thought to be
recurrent, by convention RNe are those novae that have experienced multiple recorded outbursts in the last 150 years or so.  
Pure solar accretion studies show that virtually no material is ejected and, therefore, those WDs must be
growing rapidly in mass \citep[][]{starrfield_2012_basi, starrfield_2014_aa}.

\section{Conclusions}

\begin{enumerate}

\item The amount of accreted material is an inverse function of 	the initial abundance of $^{12}$C.

\item By first accreting solar material (rather than mixed), more matter is accreted than if we assumed mixing from the beginning.  Reducing the metallicity to
values in agreement with the Magellanic Clouds, or even lower, further reduces the initial $^{12}$C abundance allowing more material to be accreted before the
TNR is initiated \citep{starrfield_1999_ac, jose_2007_aa}. 

\item Either no mixing with the WD (solar accretion) or mixing too early with the WD (MFB) results in an outburst that is less violent and 
little material (accreted plus WD) is ejected during the outburst.  This also causes the CO WD to grow in mass.  
We have shown this both by following one outburst with NOVA and repeated outbursts with MESA \citep{starrfield_2016_aa}. 

\item Multi-dimensional studies show that there is sufficient mixing during the TNR to agree with observations of the ejecta abundances
 \citep[][and references therein]{casanova_2018_aa}.  This mixing occurs via convective entrainment (dredge-up of WD outer layers into the accreted material)
  during the TNR and does not affect the total amount of accreted material since it occurs after the accretion phase of the outburst.

\item Simulations with 25\% WD and 75\% solar matter, mixed after the TNR is underway, eject only a fraction of the accreted material.  Therefore, 
the WD is {\it growing} in mass as a result of the Classical Nova phenomena (see Figure \ref{figuremassej}).

\item Simulations with 50\% WD and 50\% solar matter, mixed after the TNR is underway, ejected a larger fraction of accreted material but not as much as
was accreted.  Therefore, these simulations, with more $^{12}$C, also imply
that the WD is growing in mass as a result of the Classical Nova phenomena.  They also reached higher peak temperatures 
and ejected more material moving at higher velocities 
than those with only 25\% WD and 75\% solar matter.

\item Our simulations confirm that CO novae are overproducing $^7$Be, which decays to $^7$Li after we have ended our simulations.  
This result is in agreement with the observations of enriched $^7$Be in CN explosions, although the observed values exceed our predictions and those of others.

\item Our simulations show that the analyses of  \citet{nomoto_1982_aa} and \citep{fujimoto_1982_aa, fujimoto_1982_ab} are not supported by modern evolutionary 
or hydrodynamic simulations and that, by themselves, do not argue against the single degenerate scenario for SN Ia progenitors.

\item While we do not rule out the SSS as SN Ia progenitors, their observed numbers suggest that they
are likely to be an extremely small channel with typical CVs being a major channel.  Finally, the observations of SN Ia explosions alone suggest that there are
multiple channels for their progenitors \citep[][and references therein]{nugent_2019_aa}.

\item Our results indicate that even systems with low accretion rates, \.M $< 10^{-9}$ M$_\odot$ yr$^{-1}$, can produce CNe in which the WD is 
growing in mass toward the Chandrasekhar Limit.  In combination with the results of \citet[][done with a
different code and higher mass accretion rates]{hillman_2015_ac, hillman_2016_aa}, 
our simulations add a considerable area to the \.M - WD mass plane, where evolution to a SN Ia is possible.
It is no longer necessary to assume that the only area in which the WD grows in mass is that region designated as the ``Steady Burning'' region.  

\end{enumerate}

We acknowledge useful discussion and encouragement from M. Darnley, E. Aydi, J. Jos\'e,  M. Hernanz, A. Heger, S. Kafka,  L. Izzo, P. Molaro, M. della Valle, A. Shafter and the attendees at EWASS18, COSPAR 2018, and HEAD 2019 for  their comments.  We thank the anonymous referee, J. Jos\'e, and A. Shafter for their comments on an earlier draft which greatly improved this manuscript.
This work was supported in part by NASA under the Astrophysics Theory Program grant 14-ATP14-0007 and the U.S. DOE under Contract No. DE-FG02- 97ER41041. SS acknowledges partial support from NASA, NSF and HST grants to ASU, WRH is supported by the U.S. Department of Energy, Office of Nuclear Physics, and CEW acknowledges support from NASA and NSF.

\bibliography{references_iliadis,starrfield_master}

\end{document}